\documentclass[12pt,preprint]{aastex}
\makeatletter

\newcommand{\Rmnum}[1]{\expandafter\@slowromancap\romannumeral#1@}
\makeatother
\usepackage{amsmath,amssymb}
\slugcomment{accepted by ApJ} 
\shorttitle{The dynamics of funnel prominences}
\shortauthors{Keppens and Xia}
\begin{document}
\title{THE DYNAMICS OF FUNNEL PROMINENCES}
\author{R. Keppens\altaffilmark{1,2}, C. Xia\altaffilmark{1}}
\altaffiltext{1}{Centre for mathematical Plasma Astrophysics, Department of 
Mathematics, KU Leuven, Celestijnenlaan 200B, 3001 Leuven, Belgium}
\altaffiltext{2}{School of Astronomy and Space Science, Nanjing University, 
Nanjing 210093, China}

\begin{abstract}
We present numerical simulations in 2.5D settings where large scale prominences form in situ out of coronal condensation in magnetic dips, in close agreement with early as well as recent
reporting of `funnel prominences'. Our simulation uses full thermodynamic MHD with anisotropic thermal conduction, optically thin radiative losses, and parametrized heating as main ingredients to establish a realistic arcade configuration from chromosphere to corona. The chromospheric evaporation from especially transition region heights ultimately causes thermal instability and we witness the growth of a prominence suspended well above the transition region, continuously gaining mass and cross-sectional area. Several hours later, the condensation has grown into a structure connecting the prominence-corona transition region with the underlying transition region, and a continuous downward motion from the accumulated mass represents a drainage that matches observational findings. A more dynamic phase is found as well, with coronal rain, induced wave trains, and even a reconnection event when the core prominence plasma weighs down the fieldlines until a fluxrope gets formed. The upper part of the prominence is then trapped in a fluxrope structure, and we argue for its violent kink-unstable eruption as soon as the (ignored) length dimension would allow for ideal kink deformations.   
\end{abstract}


\section{INTRODUCTION}\label{intro}

Although prominences have fascinated solar physicists for decades, the latest IAU symposium 300 dedicated to the topic came to a sobering conclusion: `The question "How do prominences from?" is still open'~\citep{priest}. While several possibilities to achieve hundredfold denser and cooler plasma conditions in the corona are known, detailed modeling of prominence formation remains challenging. Especially the route through thermal instability~\citep{field,parker}, which requires the inclusion of full thermodynamics with radiative losses depending on density-temperature conditions, has convincingly been modeled in 1D settings~\citep{mok90,antiochos99,karpen01,Xia11,luna12,zhang13}. Multidimensional aspects are less explored, although the 1D approach can be combined with rigid 3D fields~\citep{luna12,schmit14}, and give first hints of how projection effects matter within 3D topologies. A breakthrough was made by~\cite{Xia2012ApJ}, when a 2.5D bipolar arcade was subjected to the evaporation-condensation process known to trigger thermal instability~\citep{Xia11}. There, the adopted magnetohydrodynamic (MHD) simulation, including thermodynamics, formed a true macroscopic condensation of size and accumulated weight that matches quiescent prominences. The dense plasma thereby dipped fieldlines while condensing, to form a virtually ideal MHD force-balanced state, much akin to the original analytic work by~\cite{ks}. Non-ideal effects occur primarily at the prominence corona transition region interface (PCTR), and involve the detailed interplay of anisotropic thermal conduction with heating and cooling operative. The catastrophic cooling is also at play for coronal rain, where earlier 1D models~\citep{muller05,patrick10} have been extended to 2D arcade evolutions~\citep{fang13}. 

Most prominences appear embedded in coronal cavities~\citep{gibson06}, and these can provide morphological information on possible lack of ideal MHD equilibrium conditions, signaling coronal mass ejection onset. While advanced force-free 3D models exist that focus on the complex magnetic field topology where upwardly dipped parts may host prominence material~\citep{aulanier98}, and recent zero-beta simulations provide close matches with the dynamics seen during violent prominence-loaded CMEs~\citep{kliem12}, force-free and zero-beta simulations by definition contain no prominence thermal structure, and frequently ignore gravity. Especially the latter is obviously required when wanting to model quiescent prominence dynamics, where magnetohydrostatic computations can solve for the Grad-Shafranov type equations following from 2.5D assumptions~\citep{petrie07,blokland11}. Those magnetohydrostatic conditions are the starting point for detailed prominence seismology~\citep{blokland11b}, which can extend current insights based on simple geometric models~\citep{ballester}. The step to multidimensional studies of linear waves in prominences has meanwhile been made using source-term injected plasma in a 2D potential arcade system~\citep{jaume13}, and we will here adopt the quadrupolar arcade from that study, to demonstrate the in-situ formation of prominences due to evaporation-condensation. The pre-prominence arcade already has field lines that show a dipped structure, but does not meet the requirements for fluxrope-cavity structures that are so often observed. 
The dipped arcade adopted here is more appropriate for prominences that have in early classifications as well as more recently been termed `funnel prominences'. The original terminology relates to their appearance in $H_\alpha$ filters showing an inverted cone structure, and~\cite{kleczek} uses historic data (from Menzel and Evans, taken between 1956-1961) to find that funnel shaped prominences have a compact body and little fine structure, have lifetimes in the order of hours, and are often associated with plages without sunspot groups. The recently revived funnel prominence terminology relates to similar funnel shape prominences, this time as seen by the EUV channels (especially at 171 \AA\, and 304 \AA) of SDO/AIA, and where it is suggested that these prominences and their internal dynamics may be a major player for the mass cycle in the chromosphere-corona system~\citep{liuetal}. In contrast to typical polar crown prominences embedded in coronal cavities~\citep{liu2012apj}, these are reported to form at arcade dips. The quadrupolar field topology used in our simulations also appears in early models for the original funnel prominence classification. \cite{ivanov} used a purely kinematic (cold plasma, strong field, no gravity or thermodynamics) model that hinges on the presence of two aligned dipoles introducing an X-point at some height in the solar atmosphere. When field lines reconnect at this location, the frozen-in condition external to the reconnection region could cause plasma movement acting to collect matter (lifting it upward) in a funnel-shaped region above this reconnection point. The kinematic, 2D field topology change was induced by applying time-varying dipole moment strengths. Hence, their model hinges on reconnection, and has prominence matter carried upward by Lorentz forces. Although our magnetic topology bears strong resemblance to this setup, it is of interest to already point out that we will demonstrate (1) in-situ condensation after chromospheric evaporation; (2) that the route by thermal instability does not involve reconnection (at least, not in the formation phase and at a different location in the later evolution), and (3) we do not require time-varying bottom magnetic field evolutions.

Motivated by the most recent observational results, we present here a detailed study of a simulation that follows prominence condensation for hourlong periods. In Section~\ref{numer}, we list all details on the numerical setup and discretizations adopted. Section~\ref{resu} starts with a qualitative description of the full 12 hour simulation period, to then turn to quantitative findings on mass, thermodynamics and overall topological changes. A summary and outlook, especially identifying aspects needing future attention, is given in Section~\ref{conclusion}.

\section{COMPUTATIONAL ASPECTS}\label{numer}

\subsection{Initial setup and governing equations}

The initial thermodynamic state is constructed from a 1D stratified equilibrium as follows. The variation of temperature $T$ with height $y$ is first computed from prescribing the transition region temperature $T_{\mathrm{tr}}=1.6 \times 10^5 \,{\mathrm{K}}$ at a chosen height $h_{\mathrm{tr}}=0.27 \times 10^7 \,\mathrm{m}$, and demanding a constant vertical thermal conduction flux prevailing towards increasing heights. The latter requires $\kappa(T) \frac{dT}{dy}=200 \, \mathrm{J}\mathrm{m}^{-2} \mathrm{s}^{-1}$, while below $h_{\mathrm{tr}}$ we adopt the fixed value $T_{\mathrm{b}}=10^4 \, \mathrm{K}$. From this temperature stratification, the density $\rho$ and pressure $p$ variation are determined from hydrostatic balance, under a given bottom number density value $n_{\mathrm{b}}$ (related to bottom density through $\rho_{\mathrm{b}}=1.4 m_{\mathrm{p}} n_{\mathrm{b}}$ under a fully ionized plasma with 10:1 H:He abundance) of $2.5 \times 10^{20} \,\mathrm{m}^{-3}$. To use the hydrostatic balance in the initialization, but also in the bottom boundary prescription as mentioned further on, we actually first compute a 1D pressure and density array at an arbitrarily high resolution from the ideal gas law combined with the discrete formula
\begin{equation}
\frac{p_j-p_{j-1}}{\Delta y} = \frac{1}{4} \left(g_j+g_{j-1}\right) \left(\frac{p_j}{T_j}+\rho_{j-1}\right) \,,\label{hdstatic}
\end{equation}
where $g_j$ indicates the local solar gravity value $g(y)=-274 \frac{R_\odot^2}{(R_\odot+y)^2}\, \mathrm{m}\,\mathrm{s}^{-2}$. The initial density and pressure variation on the AMR grid is then found from a direct interpolation within this array. The velocity is set to zero throughout, and the magnetic field topology is taken as
\begin{eqnarray}
B_x & = & + B_{p0} \cos\left(\frac{\pi x}{2 L_0}\right) e^{-\frac{\pi y}{2 L_0}} - B_{p0} \cos\left(\frac{3 \pi x}{2 L_0}\right) e^{-\frac{3 \pi y}{2 L_0}} \,,\nonumber\\
B_y & = & - B_{p0} \sin\left(\frac{\pi x}{2 L_0}\right) e^{-\frac{\pi y}{2 L_0}} + B_{p0} \sin\left(\frac{3 \pi x}{2 L_0}\right) e^{-\frac{3 \pi y}{2 L_0}} \,,\nonumber\\
B_z & = & B_{z0} \,.
\end{eqnarray}
We set $L_0= 5 \times 10^7 \,\mathrm{m}$, and fix the field completely by requiring that the total field strength at a specific location $B(x=0,y=2\,L_0/\pi)$ equals $4\times 10^{-4}\, \mathrm{T}$, while the local angle $\alpha$ between the $(x,y)$ plane and the field there is fixed at $\alpha(x=0,y=2\,L_0/\pi)=\pi/4$. Note that these requirements determine $B_{z0}$ and $B_{p0}$ uniquely. The potential field given above is inspired by a similar quadrupolar field configuration adopted in~\cite{jaume13} to study linear MHD wave motions in prominences.

This initial condition represents an ideal MHD, stratified equilibrium, but we actually simulate the MHD equations extended with non-ideal effects including optically thin radiative losses $Q$, anisotropic (field-aligned) thermal conduction, and a parametrized heating function $H$. These terms appear in the evolution equation for the total energy as follows
\begin{equation}
 \frac{\partial E}{\partial t}+\nabla\cdot\left(E\mathbf{v}+p_{tot}
  \mathbf{v}-\mathbf{BB}\cdot\mathbf{v}\right)=\rho\mathbf{g}\cdot
  \mathbf{v}+\nabla\cdot\left(\boldsymbol{\kappa}\cdot\nabla T\right)-Q+H \,, \label{energy}
\end{equation}
where $E=p/(\gamma-1)+\rho v^2/2+B^2/2$ is the total energy density (we adopt $\gamma=5/3$ and set permeability $\mu_0=1$) and $p_{tot}\equiv p+B^2/2$ the total
pressure. For thermal conduction, we use a pure field-aligned dependence quantified by $\kappa_\parallel=10^{-11} T^{5/2} \, \mathrm{J}\mathrm{m}^{-1}\mathrm{s}^{-1}\mathrm{K}^{-1}$. The optically thin cooling uses a tabulated temperature dependence $\Lambda(T)$ and through $Q\propto n_H^2 \Lambda(T)$ scales with the squared hydrogen number density. The cooling table has been used in our earlier 1D and 2.5D models and contains updated data for solar coronal plasma conditions as provided by~\cite{colganetal08}. The table has a lower cut-off at 10000 K, and to evaluate the radiative loss term, we use the exact integration method as introduced by~\cite{Townsend} and intercompared to standard (semi-)implicit evaluations in AMR settings in~\cite{vanmarlekeppens}. The anisotropic thermal conduction is treated using explicit subcycling on the source update. The parametrized heating term has the following prescription
\begin{eqnarray}
H & = & H_{\mathrm{bg}}+H_{\mathrm{lh}}\,, \nonumber \\
H_{\mathrm{bg}}(y) & =  & H_0 \exp(-y/L_{\mathrm{bg}}) \,,\nonumber \\
H_{\mathrm{lh}}(x,y,t) & = & H_1 \,R(t)\, C(y) \,\left[\exp(-\frac{(x-x_r)^2}{\sigma^2}) + \exp(-\frac{(x-x_l)^2}{\sigma^2})\right] \,,\nonumber \\
C(y) & = & \begin{cases} 1 & \text{if $y < y_h$}, \\
                         \exp(-(y-y_h)^2/\lambda_h) & \text{if $y \geq y_h$} \,. \end{cases} 
\label{heateq}
\end{eqnarray}
The above formulae distinguish between a background heating $H_{\mathrm{bg}}$ for which the amplitude is fixed at $H_0=3 \times 10^{-5} \,\mathrm{J}\mathrm{m}^{-3}\mathrm{s}^{-1}$ and the scale height is $L_{\mathrm{bg}}=5 L_{\mathrm{unit}}$. The localized heating $H_{\mathrm{lh}}$ has a ramp function $R(t)$ that varies linearly between zero and one, from a given start time and within a given ramp duration. The amplitude for this localized heating is set to $H_1= 2 \times 10^{-3} \,\mathrm{J}\mathrm{m}^{-3}\mathrm{s}^{-1}$, two orders of magnitude above the background rate. The other parameters control the overall heating deposition throughout the simulated domain. We used a domain of size $[-5 L_{\mathrm{unit}}, 5 L_{\mathrm{unit}}] \times [0, 8 L_{\mathrm{unit}}]$ and took $x_l=-x_r=-4.2 L_{\mathrm{unit}}$, $y_h=0.4 L_{\mathrm{unit}}$ and $\sigma^2=0.2 L_{\mathrm{unit}}^2$ with $\lambda_h=0.25 L_{\mathrm{unit}}^2$. The simulation is performed in dimensionless fashion, where our unit of length is $L_{\mathrm{unit}}=10^{7} \mathrm{m}$, the density unit is $\rho_{\mathrm{unit}}=1.4 m_{\mathrm{p}} n_{\mathrm{unit}}= 2.3417 \, \times 10^{-12} \,\mathrm{kg}\, \mathrm{m}^{-3}$ and pressure unit $p_{\mathrm{unit}}=0.03175 \,\mathrm{J}\,\mathrm{m}^{-3}$. This normalization implies a magnetic field unit of about $2\times 10^{-4} \mathrm{T}$ and time unit of 85.87 seconds.

The initial state is first evolved to thermodynamically adjust to the combined effects of background heating $H_{\mathrm{bg}}$, conduction and radiative losses. That part of our simulation follows the establishment of an overall heated arcade that connects chromospheric, transition region to coronal plasma as simulated for 1.19 hours (50 dimensionless time units) under identical boundary and discretization settings as adopted for our main simulation. In all results shown further on, our time $t=0$ is taken as this endstate, which in particular is no longer a mere potential field, and has as a result of thermodynamic adjustments created localized current distributions and finite Lorentz forces, in response to the heat sinks and sources. In this state, the ratio of magnetic to thermal, and magnetic to kinetic energy, all computed over the entire domain $V= 80 L_{\mathrm{unit}}^2$, is 1.41 and 8114.51, respectively. The average plasma beta is 0.538, while the mean temperature over the domain is $\bar{T}=\frac{1}{V}\iint T \,dx\,dy = 2.28 \times 10^6 \mathrm{K}$.  

\subsection{Discretization, AMR, and boundary treatments}

For numerically advancing the governing PDEs, we use a three-step Runge-Kutta type scheme (details are given in~\cite{keppensporth} and references therein) and a third-order accurate limited reconstruction introduced by~\cite{cada} to go from cell center to cell edge variable evaluation as needed for flux computations. For the fluxes, we employ a suitably mixed prescription between a diffusive TVDLF and contact-resolving HLLC scheme, as introduced in relativistic hydro settings by~\cite{zak2008} but here used for newtonian MHD settings. Control on the magnetic field monopole discretization errors is effected by using a diffusive approach, intercompared to other source term treatments in AMR settings in~\cite{Keppens03}. As stated before, the source terms are treated in a variety of ways, with a split strategy for this corrective monopole diffusion, an unsplit explicit addition for gravity and heating, while anisotropic thermal conduction uses explicit subcycling and radiative losses use an exact integration approach. Overall, we use a Courant parameter of 0.9, while diffusive terms introduce a similar fractional restriction on the time step of 0.4. The simulation uses a base grid of $120\times 120$ grid points, but activates automated mesh refinement based on a mixed evaluation of weighted discrete second derivates~\citep{lohner,Kepp12}, involving density and both poloidal magnetic field components in a $0.6:0.2:0.2$ ratio. We allow for 3 levels, reaching effective mesh size of $480\times 480$, with smallest cell widths $\Delta x\approx 208 \, \mathrm{km}$ and heights $\Delta y \approx 167 \,\mathrm{km}$.
 We ensure that the bottom region up to $y=0.05 L_{\mathrm{unit}}$ is always treated at this maximal resolution.

As boundary conditions, we make use of ghost cells, which prescribe cell center values in 2 grid layers exterior to the domain. At left and right physical boundary, we use symmetric conditions on density, energy, $y$- and $z$-momentum components as well as on $B_y$, $B_z$. Asymmetric conditions, ensuring zero face values, are adopted for $v_x$ and $B_x$. At bottom, we fill the ghost cell primitive variables $(\rho, \mathbf{v}, p, \mathbf{B})$ by asymmetry on all velocity components, fixing the analytic potential expression for $\mathbf{B}$, and fixing the gravitationally stratified density and pressure variation exploited for the initial state. At the top, we similarly enforce no-flow through conditions, and use a discrete pressure-density extrapolation from the top layer pressure, ensuring a maximal ghost layer temperature $T_{\mathrm{top}}= 2\times 10^{6} \mathrm{K}$ through the gravitational field. For the magnetic field, we use a second order one-sided zero-gradient condition on $B_x$, fix $B_z=B_{z0}$ and determine $B_y$ in the ghost cells from a centered difference evaluation of $\nabla\cdot\mathbf{B}=0$.

As already mentioned, we restart ($t=0$) from a thermodynamically relaxed state, from which point on we now activate the local heating $H_{\mathrm{lh}}$. Its ramp $R(t)$ duration is set to 500 seconds, and we further on describe what happens when following this evolution for about 12 hours (500 code time units). 

\section{RESULTS}\label{resu}

\subsection{Prominence birth, growth and evolution}\label{frbirth}

\begin{figure}
\centering
\includegraphics[width=4.in]{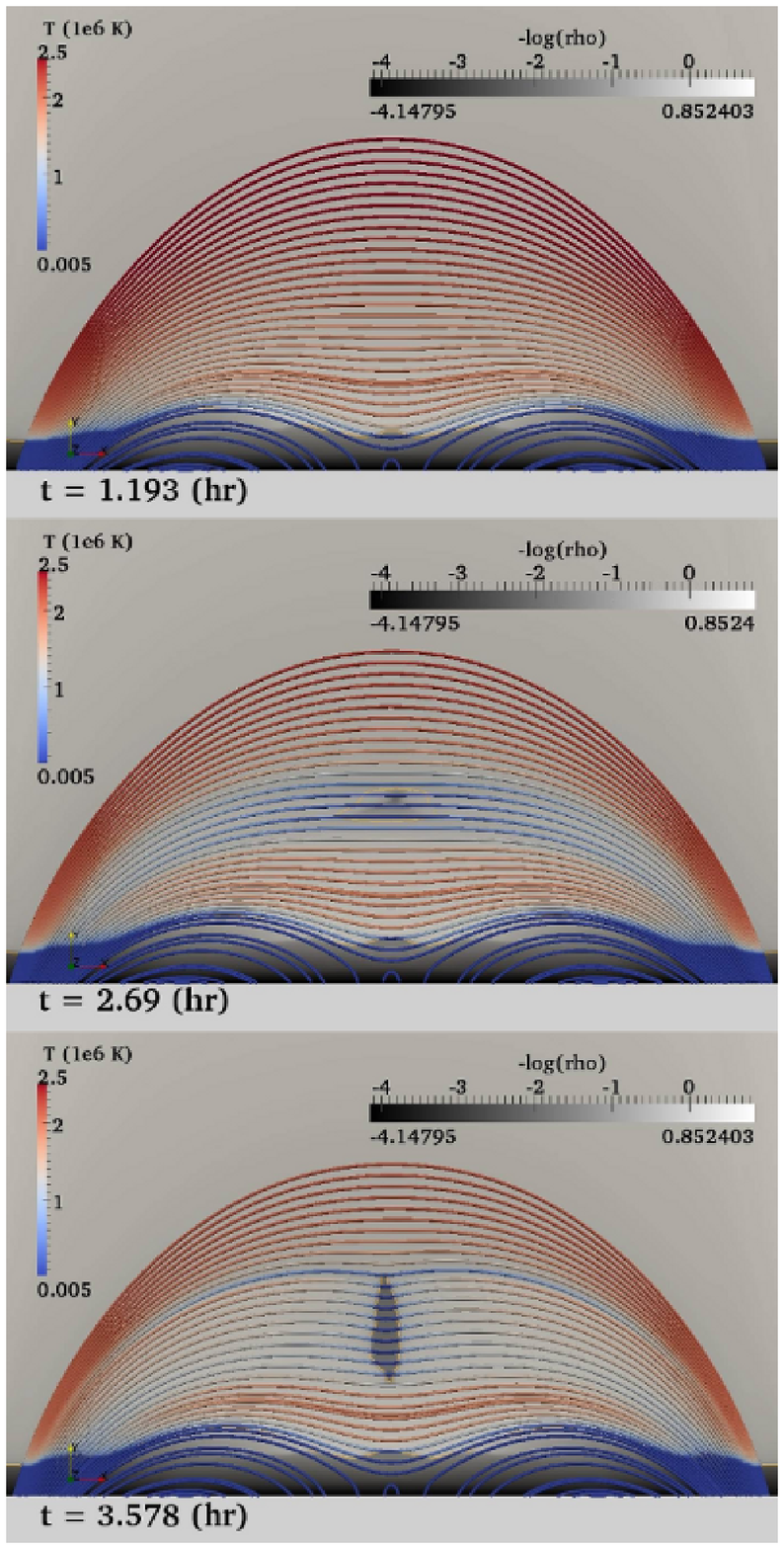}
\caption{
Snapshots before (top) and during (bottom) prominence growth, and at condensation onset (middle). Greyscale for logarithmically stretched density, fieldlines colored with temperature. A thin orange line indicates the $T=100000\, \mathrm{K}$ isocontour, marking transition region and PCTR. (A color version of this figure is available in the online journal.)
}
\label{fevolve1}
\end{figure}

Since the additional heating is affecting only a part of the quadrupolar arcade system, and preferentially heats chromospheric to transition region heights (this is here fully parametrically prescribed by the choices for $y_h$ and $\lambda_h$ in formulae~(\ref{heateq})), not much happens for quite a while throughout most of the domain. The transition region moves slightly upwards in the heated areas around both footpoints $x_r=4.2 \times 10^7 \mathrm{m}$, $x_l=-x_r$, which connect shallow-dipped overarching arcade fieldlines. A continuous chromospheric evaporation gradually increases the density in the heated loop segments, and similarly decreases their temperature. This continues for about 2.69 hours, at which point the thermal instability process begins and runaway catastrophic cooling occurs in a bundle of overarching fieldlines. The roughly 4-5 Mm width of this bundle corresponds to the choice for the heating parameter $\sigma^2$. The density and temperature locally change by two orders of magnitude, and a macroscopic prominence emerges, continuously growing in area and mass. By a time 3.578 hours after the added heating was switched on, a central 3 Mm wide and 12.5 Mm tall prominence is situated at an altitude of about 20 Mm, still well above the transition region. Figure~\ref{fevolve1} shows the magnetic field, temperature and density distribution through the relevant lower domain part. The density greyscale plots minus the logarithm of dimensionalized density values, i.e. its zero value relates to our density unit $\rho_{\mathrm{unit}}$. A movie of the complete time evolution, using the same visualization, is provided in online material. The field lines are colored by the temperature variation with cooler regions in blue, and a thin orange line indicates the $10^5 \, \mathrm{K}$ contour. For all times before the instability onset, this latter contour gives a clear proxy to locate the transition region (TR), as also seen in the logarithmically stretched greyscale used for the density variation. As soon as the prominence appears, it similarly locates the prominence corona transition region (PCTR). The times shown in Fig.~\ref{fevolve1} (1) represent a view at 1.193 hours, typical for the entire first 2.69 hour period; (2) the middle panel is showing the sudden central condensation forming; and (3) the snapshot for 3.578 hours is representative for the first quiescent state that lasts up to about $t=6$ hours in our simulation. In this quiescent state, the prominence continues its growth both in width and height, and ultimately forms a structure that connects down to the underlying transition region (joining the PCTR and TR temperature contour as mentioned above). While the prominence grows, it enhances the dips of the fieldlines on which it resides, but overall establishes force-balanced conditions. This phase is largely identical to the findings reported in~\cite{Xia2012ApJ}, where both horizontal and vertical force analysis was performed on a macroscopic prominence forming on top of a pure bipolar non-linear force-free arcade. The main differences thus far relate to (1) field topology, with our quadrupolar arcade already having dipped fieldlines prior to the prominence onset, (2) the heating function parametrization, and (3) the fact that we here followed the evolution for much longer times (\cite{Xia2012ApJ} showed 2.5D prominence growth for about 1 hour). Note that we now simulate the entire arcade system, while \cite{Xia2012ApJ} controlled the central formation of a normal polarity, Kippenhahn-Schl\"uter~\citep{ks} type structure, by imposing mirror symmetry as a boundary condition. In the quadrupolar case, a slight asymmetry is present from the outset, which is entirely due to accumulated discretization and round-off errors in our parallel, grid-adaptive evolution. 
As a result, we find the first condensation forming at $x=0.104167$ Mm, slightly off-center, at a height of $y=21.0833$ Mm. When growing into a macroscopic prominence, further asymmetries can be detected, with indications of wave motions occuring throughout the increasing prominence body. In addition, a slow bodily sideways movement of the prominence happens, and we will quantify this statement further on. As soon as the prominence forms, also a downward trend is seen, and the mass-loaded portions of the arcade fieldlines become increasingly upwardly dipped. By visual inspection of the animated views, it can be seen that especially the lower part of the prominence body, once formed, shows an inverted funnel shape. Combined with its appearance within a dipped arcade topology, we can safely associate our model with the funnel prominences reported by observations.

\begin{figure}
\centering
\includegraphics[width=4.in]{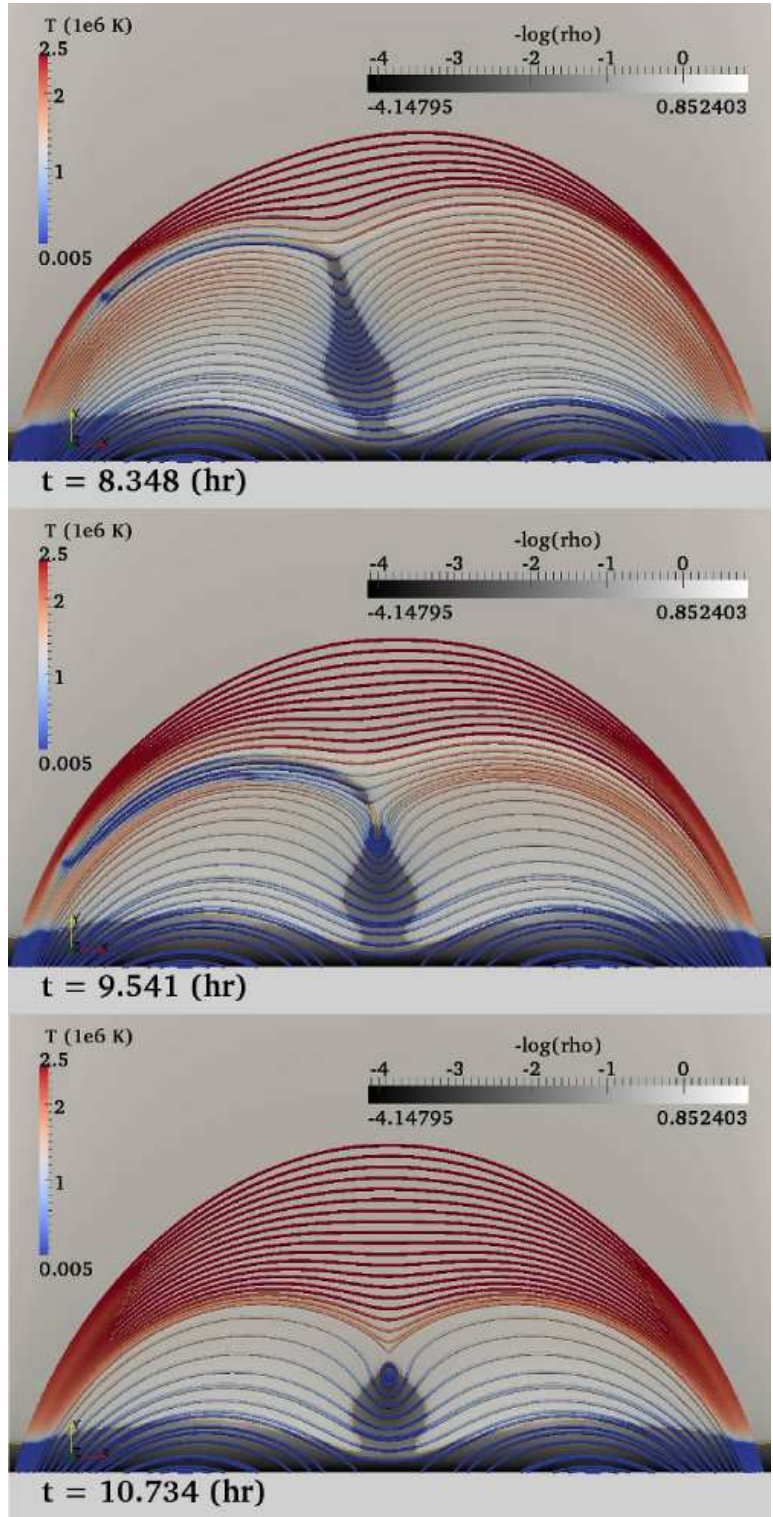}
\caption{
As Figure~\ref{fevolve1}, for selected times in the more dynamic phases. Top: coronal rain onset. Middle: fluxrope formation. Bottom: fluxrope embedded prominence phase.
(A color version of this figure is available in the online journal.)
}
\label{fevolve2}
\end{figure}

From about 6 hours into the simulation, the prominence gets seemingly connected to the transition region, and we qualitatively distinguish a second, more dynamic evolution phase between 6 and 9.5 hours. During this phase, we witness the prominence body orienting itself (left) slanted to the vertical. Ultimately, the very top part of the prominence body `spills over' and causes a transient coronal rain event with fragments of the prominence body falling down the arched fieldlines towards the transition region. Figure~\ref{fevolve2} shows three more snapshots from our simulation, with the top panel at time $t=8.348$ hours demonstrating the rain event, and the middle panel at $t=9.541$ hours marking the end of this second phase. In fact, at that time the combined effect of accumulated mass within the central prominence body, augmented with impulsive dynamics due to the transient coronal rain, causes a (numerical) reconnection event that forms a finite-sized fluxrope within the upper part of the prominence body. For the remaining times, from $t=9.5$ up to time $t\approx 12$ hours, this fluxrope remains and co-evolves with the ever descending prominence structure. The snapshot in Fig.~\ref{fevolve2} at $t=10.734$ hours shows how the fluxrope is seen in projected poloidal fieldlines. 
It is interesting to note that with respect to the dominant bipolar background field variation, the prominence (both in the arcade and fluxrope) has a normal polarity, but due to the quadrupolar underlying field, it is more appropriate to classify it as an inverse polarity prominence with respect to the PIL underneath.

As evident from the above discussion, we thus identify up to three phases in the prominence evolution: quiescent growth, dynamic phase and fluxrope stage. During all three, the main prominence is always resident in the dipped parts of an overall arcade system, as also suggested by the funnel prominences identified in the recent observations~\citep{liuetal}. In the remainder of this paper, we will systematically quantify various aspects.

\subsection{Thermodynamical evolution and energetics}

\begin{figure}
\centering
\includegraphics[width=\textwidth]{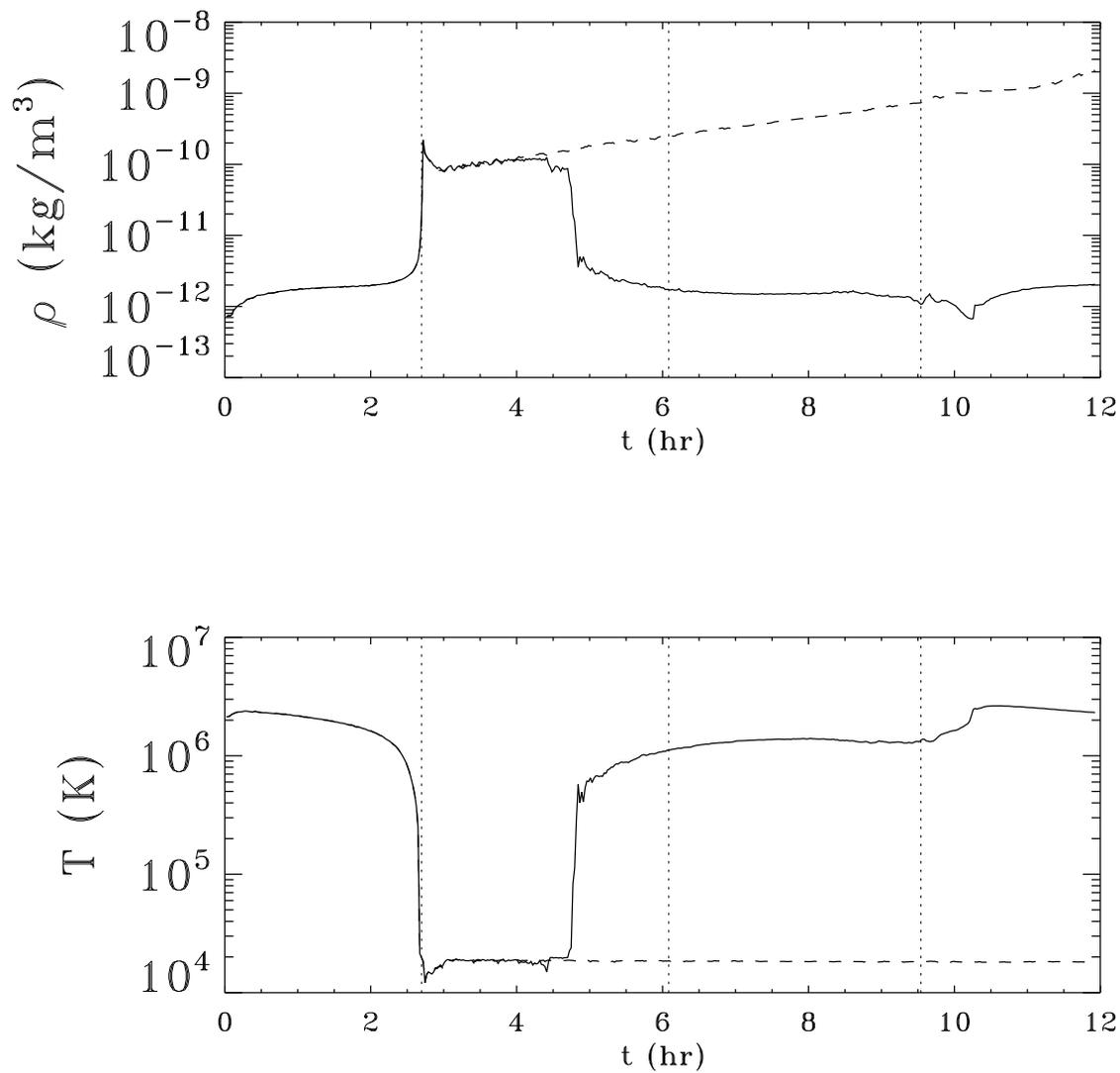}
\caption{
Evolution of the density in $\mathrm{kg}\,\mathrm{m}^{-3}$ (top) and temperature in $\mathrm{K}$ (bottom) at the near-central location of prominence onset (solid lines). The dashed line gives the same information for the instantaneous position of the prominence centre of mass. Dotted vertical lines mark condensation onset (at $t\approx 2.69$ hours), the time when PCTR joins the TR at $t\approx 6$ hours, and the time of flux rope formation ($t\approx 9.5$ hours).
}
\label{flocal}
\end{figure}

In order to quantify the prominence evolution, we can start by showing the time history of density and temperature in a specific location. Since the prominence in essence starts at the location $x=0.104167$ Mm, $y=21.0833$ Mm, an obvious choice is to plot these for this location during the entire 12 hour period. In Fig.~\ref{flocal}, this is shown as a solid line. This confirms the onset of thermal instability as already found in earlier 1D rigid fieldline models (e.g.~\cite{Xia11} and references therein) serving as trigger to catastrophic cooling. The solid lines in Fig.~\ref{flocal} seem to indicate a sudden changeover back to coronal density and temperature conditions at $t\approx 4.7$ hours. At this time, the prominence body, as a result of the systematic sideways motion mentioned earlier, moves to the left of the chosen location. In order to quantify thermodynamic and other evolutions for the prominence as a whole, we need to resort to a more consistent way to identify `prominence' matter while it moves. We therefore introduce masks to distinguish coronal versus prominence material at all times. Starting with the corona, we explained earlier that the $T=100000$ K contour serves as a good proxy for locating transition regions, so our coronal mask flags all cells from the volume $V$ where $T>100000$ K. As the prominence forms, its interior is thereby excluded due to the establishment of the PCTR. To locate the prominence matter, we adopt two seperate masks to distinguish prominence and core prominence matter. The former identifies all cells where $T< 100000$ K, while the density exceeds $\rho> 10 \rho_{\mathrm{unit}} = 2.341 \times 10^{-11} \,\mathrm{kg}\,\mathrm{m}^{-3}$, with simultaneously $|x| < 2 \,L_{\mathrm{unit}}$ and $0.6 L_{\mathrm{unit}} < y < 5 L_{\mathrm{unit}}$. The pure geometric part of our filter is an ad hoc measure to exclude denser, cool plasma below the transition region, as well as coronal rain aspects in the more dynamic phase. Visual inspection of an animation including the snapshots shown in Figs.~\ref{fevolve1}-\ref{fevolve2} show that this indeed correctly encompasses the prominence body at all times. A second filter for identifying the core prominence plasma selects the most dense prominence plasma only, with $\rho> 100 \rho_{\mathrm{unit}} = 2.341 \times 10^{-10} \,\mathrm{kg}\,\mathrm{m}^{-3}$. Due to the fact that the prominence body actually connects with the TR from about $t=6$ onwards, this core prominence filter takes additional criteria as $T< 100000$ K while $|x| < 2 \,L_{\mathrm{unit}}$ and $0.4 L_{\mathrm{unit}} < y < 5 L_{\mathrm{unit}}$. Note in particular the somewhat lower value for the height selection. 

Armed with these three masks, we quantify especially the instantaneous centre of mass of the prominence body by computing
\begin{eqnarray}
\bar{x}_p(t) & = & \frac{\iint_{\mathrm{mask}_p} x \rho \,dx\,dy }{ \iint_{\mathrm{mask}_p} \rho \,dx\,dy } \,,\nonumber \\
\bar{y}_p(t) & = & \frac{\iint_{\mathrm{mask}_p} y \rho \,dx\,dy }{ \iint_{\mathrm{mask}_p} \rho \,dx\,dy } \,.
\end{eqnarray}
In further figures, the path taken by this centre of mass is shown, visualizing the sideways swaying of the prominence body discussed before. This centre of mass can only meaningfully be quantified from about $t=2.7$ hours, and the dashed lines in Fig.~\ref{flocal} correspond to the local instantaneous value of density and temperature as quantified in this position $(\bar{x}_p(t),\bar{y}_p(t))$. This shows nice correspondence with the solid line for the specific location where the prominence happens to form, while showing that the central density continues to increase for the entire period up to $t=12$ hours. The dotted vertical lines mark the regimes we introduced earlier, so still during the quiescent phase $t\in [2.7,6]$ hours, the sideways slanting of the prominence body moves it away from its point of formation. The core temperature and density evolution do not show a dramatic changeover though when the prominence PCTR joins the TR ($t\approx 6$ hour) or when the fluxrope pinches off ($t\approx 9.6$ hours). The density is seen to exceed the core mask value of $2.341 \times 10^{-10} \,\mathrm{kg}\,\mathrm{m}^{-3}$ roughly from about 6 hours as well. By coincidence, this is virtually simultaneous with the moment when the PCTR joins the TR.

\begin{figure}
\centering
\includegraphics[width=0.49\textwidth]{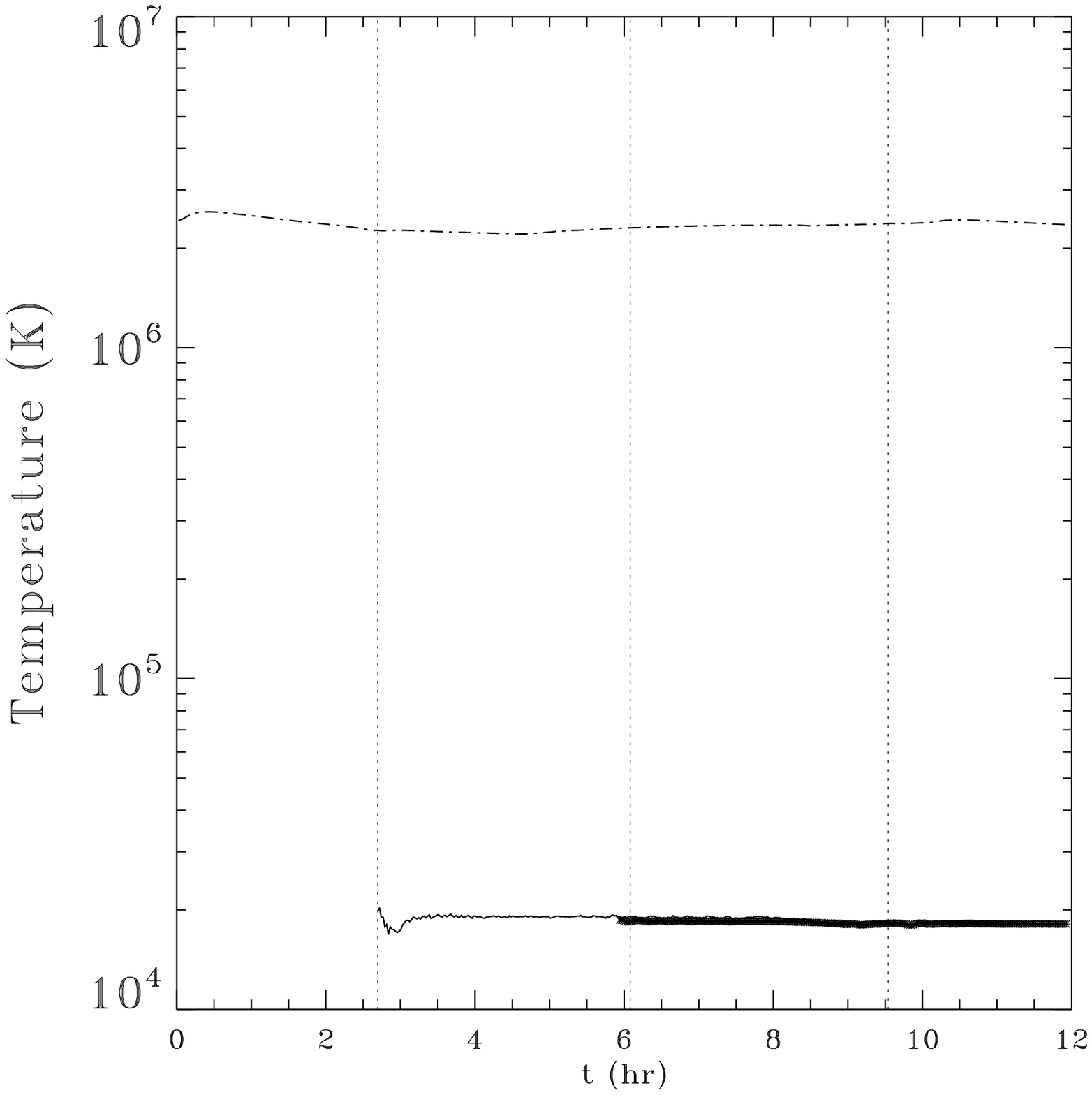}
\includegraphics[width=0.49\textwidth]{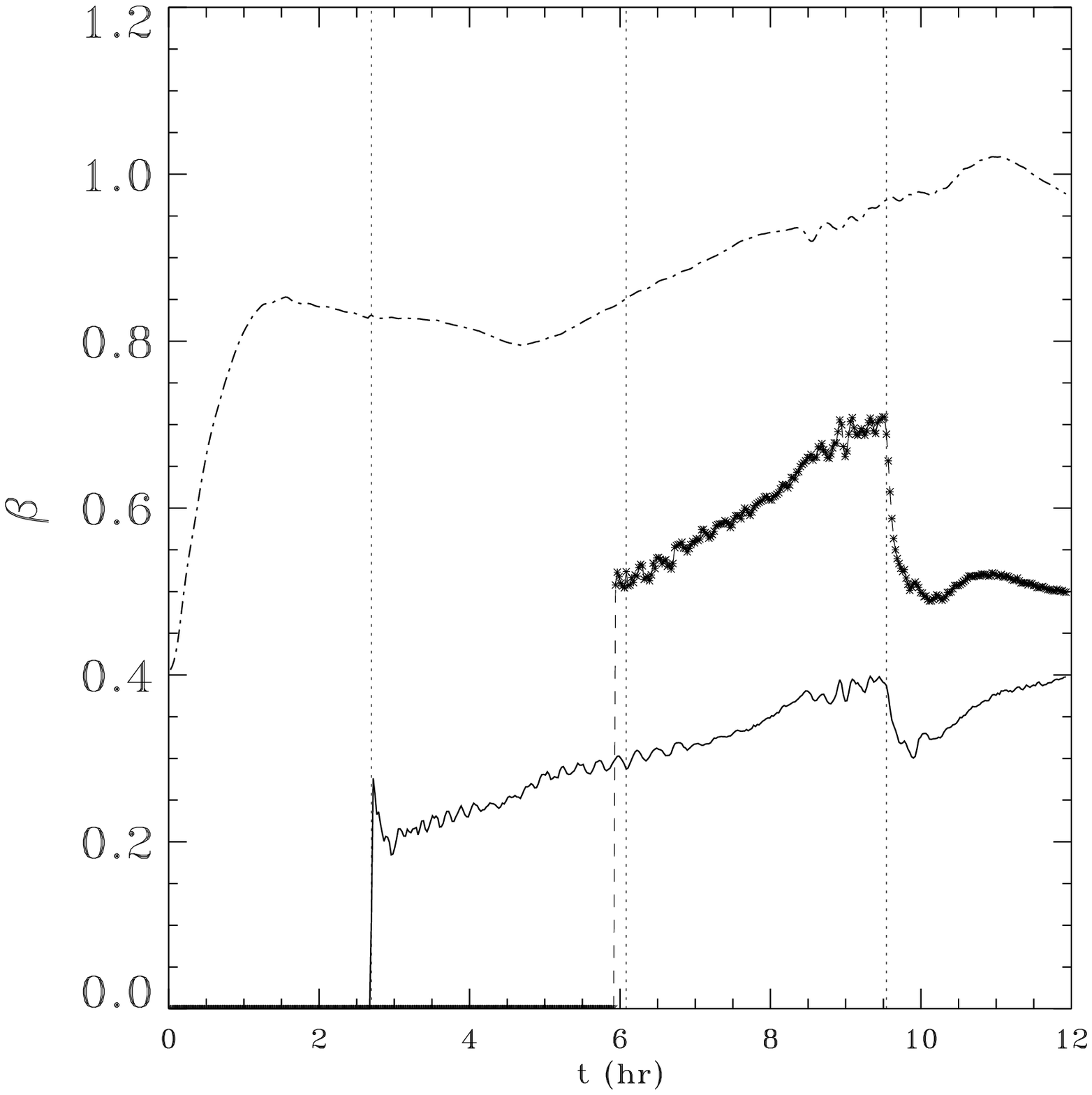}
\caption{
Evolution of the mean temperature (left) and plasma beta (right). Dash-dotted line gives the value for the corona, solid line for the prominence, and asterisk symbols (connected by dashed line) are used for the core prominence region. These three regions are identified at all times by masks as introduced in the text.
}
\label{fdriv}
\end{figure}

The three masks for corona, prominence and core prominence can then be used to quantify the average temperature, plasma beta, energetics, etc, during the evolution. For temperature and plasma beta parameters, this is shown in Figure~\ref{fdriv}. Both plots use dash-dotted lines for the coronal part, solid lines for the prominence, while the dashed line connects the asterisk symbols used for the core prominence region. As seen in the left temperature panel, the temperature conditions in the prominence interior do not vary much, staying just below 20000 K for the entire period. The two orders of magnitude temperature contrast with the coronal surroundings is in accord with known properties of prominence matter. The plasma beta evolution at right in Fig.~\ref{fdriv} shows the response to the added heating for the coronal plasma, with an overall rise in beta consistent with the increased density as a result of chromospheric evaporation. Most of the time, the coronal part has an average $\beta\approx 0.8$. In the prominence, we find that the average beta varies between $0.2-0.4$ using the first of our masks, while the core prominence material has beta varying between $0.5-0.7$, consistent with its higher density. When quantifying the plasma beta in this way, one notices that the time of fluxrope formation causes a clear drop in the average prominence beta, as influenced by combined magnetic field and thermodynamic readjustments following the reconnection event. These values for the prominence beta compare favorably with estimated beta ranges reported recently by \cite{Hillier2012}. These authors mentioned $\beta=0.47-1.13$ for reasonable $\gamma$ values, using classical fluid dynamical arguments on the observed rising plumes in quiescent prominences. 

\begin{figure}
\includegraphics[width=\textwidth]{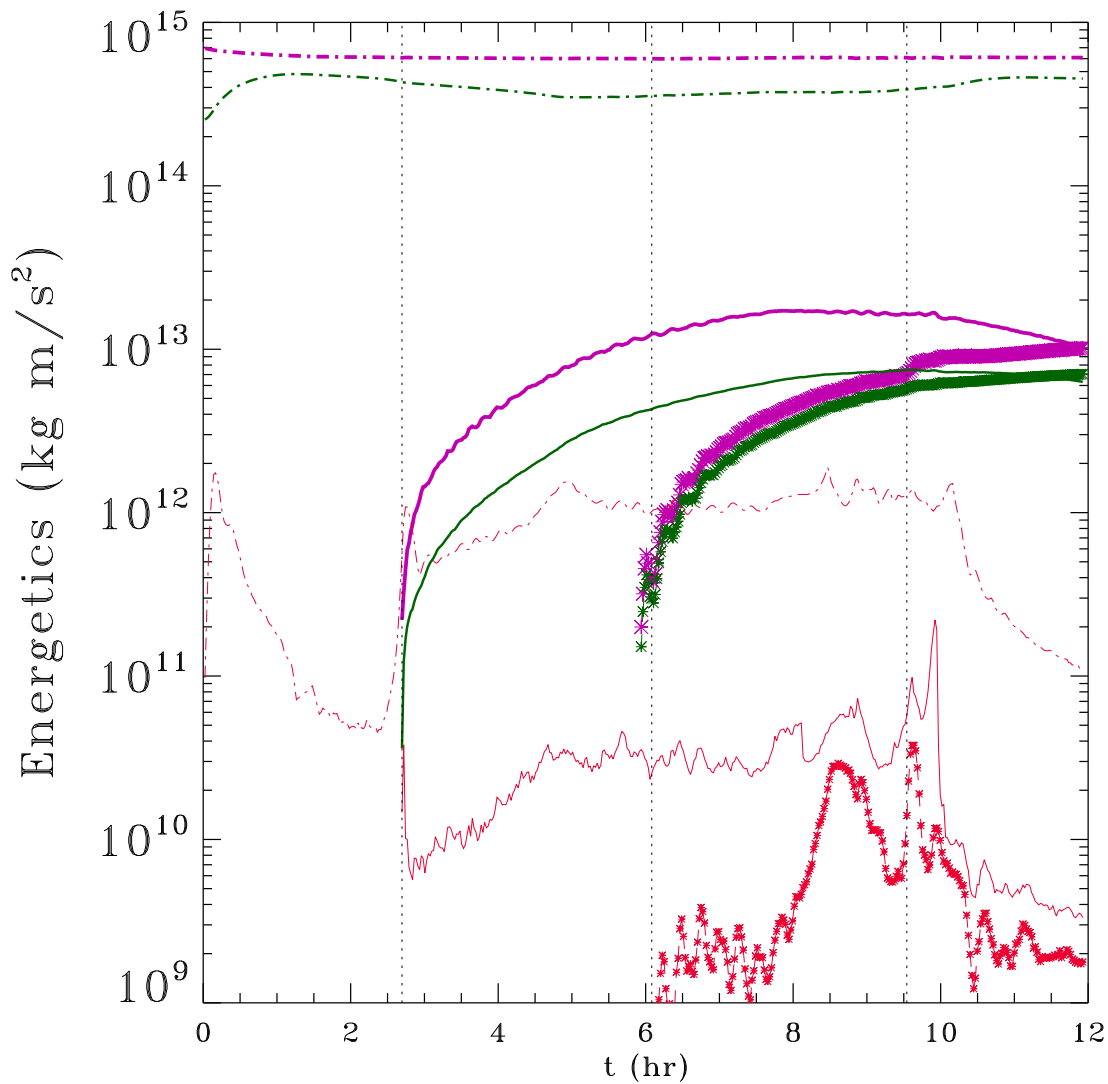}
\caption{
Using the same mask procedure used for Fig.~\ref{fdriv}, this figure quantifies the energetics as integrated over the poloidal area for the corona (dash-dotted), prominence (solid) and core prominence (connected asterisk symbols). For each region, we systematically show magnetic (purple, thickest lines), thermal (green, intermediate thickness) and kinetic energy content (red, thin lines), which retain their relative importance at all times: magnetically and thermally dominated, with a minor contribution from kinetic energy.
(A color version of this figure is available in the online journal.)
}
\label{fenergy}
\end{figure}

Figure~\ref{fenergy} collects further quantitative info on the division between magnetic, thermal and kinetic energy for all three regions identified by our masks. It is thereby seen that at all times, both in the corona and in the prominence proper, most energy is stored magnetically, followed by a thermal component. The kinetic energy is always smaller by several orders of magnitude, although it clearly displays the response to the ramp-up, to the establishment of the prominence body, and the fluxrope changeover. The oscillatory signals seen on the kinetic energy evolution for the prominence in the quiescent phase can tentatively be matched with wave motions, superposed on the sideways bodily displacements. The (slight) variation in the coronal thermal energy content also mimicks the in-situ formation and evolution of the prominence as a whole.

\begin{figure}
\centering
\includegraphics[width=0.49\textwidth]{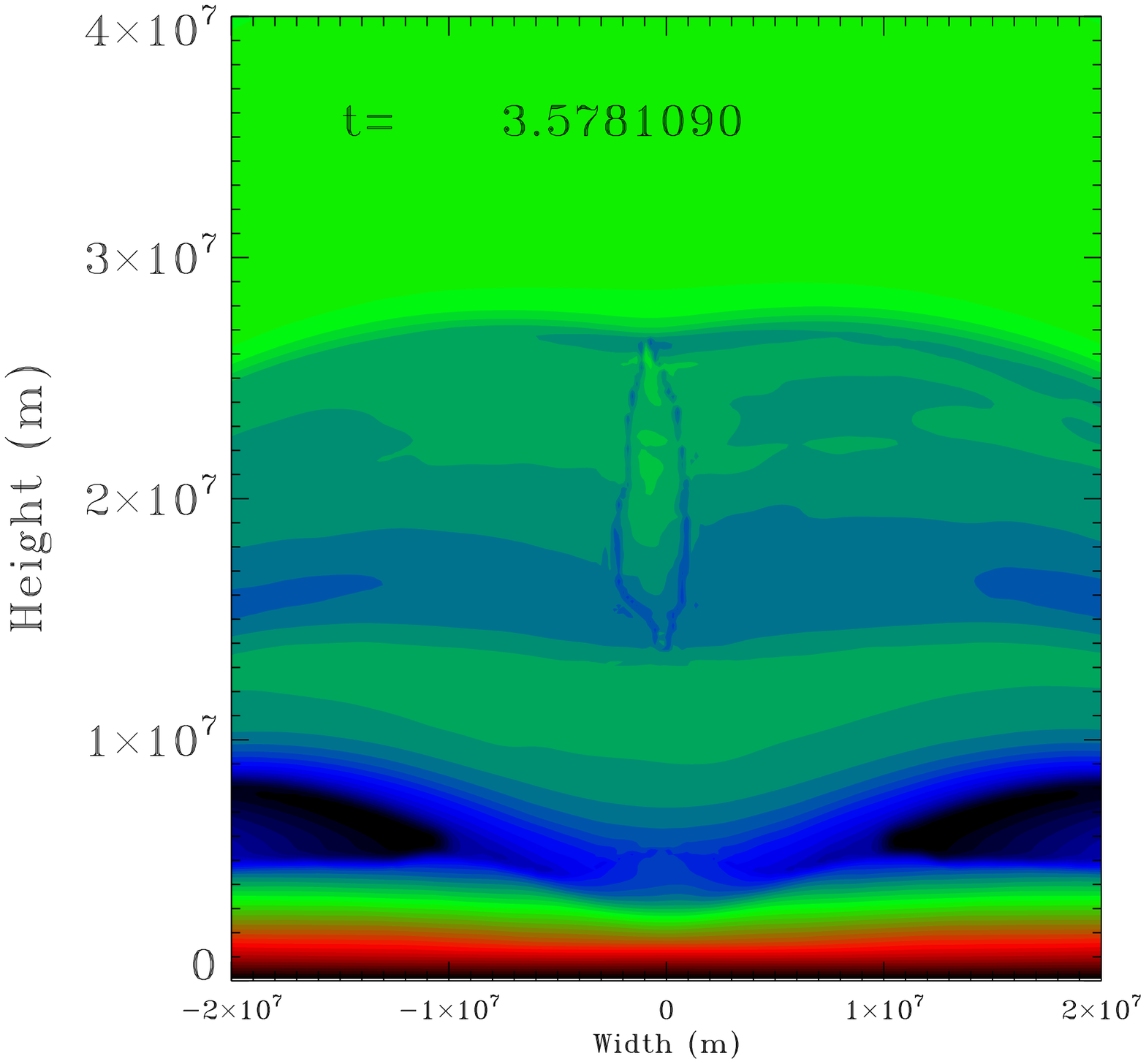}
\includegraphics[width=0.49\textwidth]{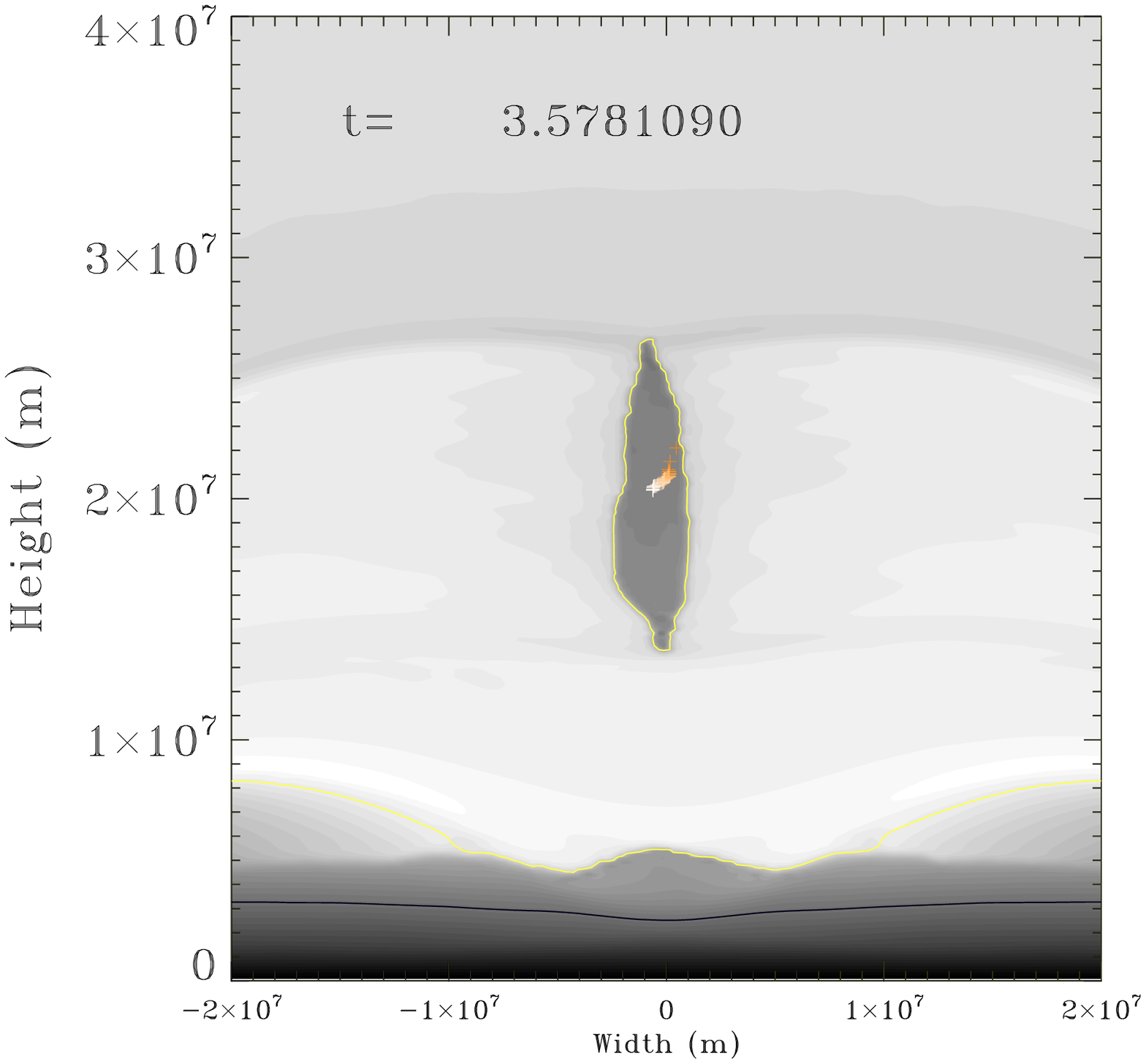}
\includegraphics[width=0.49\textwidth]{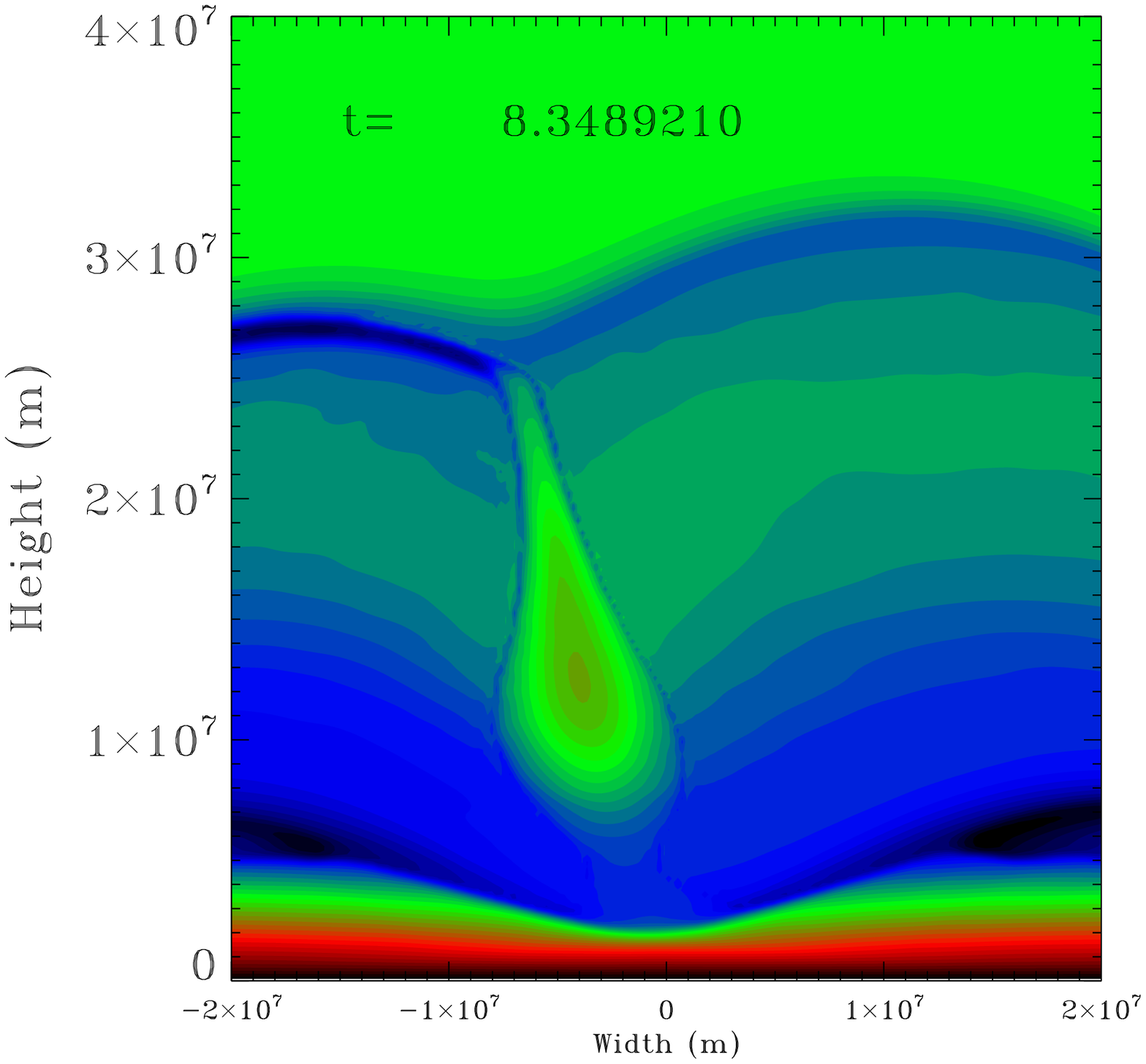}
\includegraphics[width=0.49\textwidth]{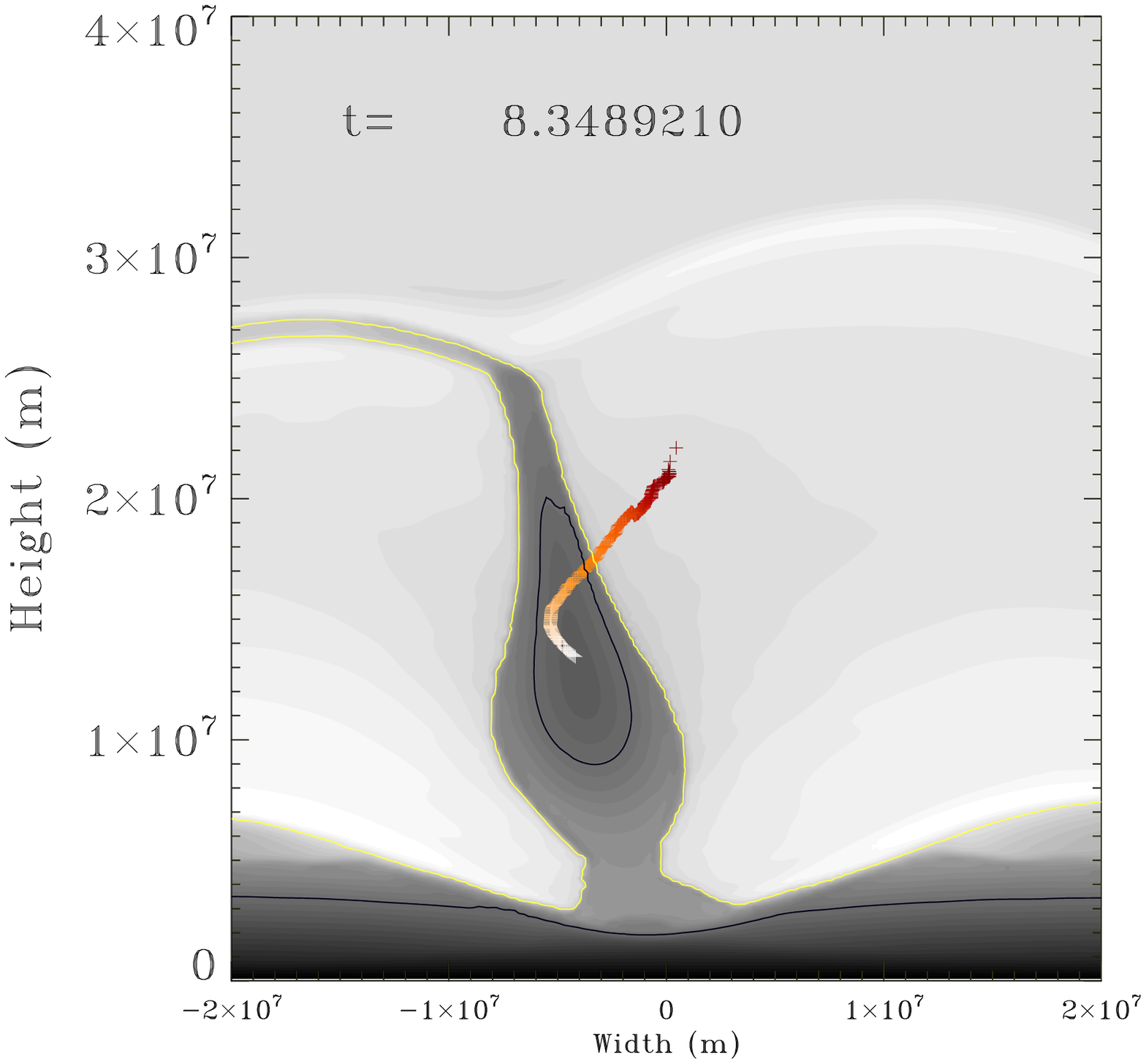}
\caption{
Zoomed views on the central regions, at two times (in hours) as indicated. Left panels show gas pressure distribution and density variation is shown at right. In the right panels, the yellow contour indicates PCTR and TR, a black contour marks the core prominence density value of $2.341 \times 10^{-10}\,\mathrm{kg}\,\mathrm{m}^{-3}$. Also in the right panels: cross-symbols indicating the path traced by the prominence center of mass, from onset to the time shown, with red to white indicating this time variation.
(A color version of this figure is available in the online journal.)
}
\label{fevolvea}
\end{figure}

To better show the prominence evolution and the relation to the masks employed to identify core prominence matter, Figure~\ref{fevolvea} only plots the central $4 \times 10^7 \mathrm{m}$ by $4 \times 10^7 \mathrm{m}$ at selected times, showing both pressure (left) and density (right) maps. In the right density frames, the yellow contour again identifies the $T=100000$ K isocontour as our TR proxy, while the black contour (only present in the bottom right panel) shows the contour selecting the core prominence region. This latter also has an isocontour below the TR proxy, justifying the need for the extra geometric filter on the masks discussed. In the right panels, we also indicate with colored cross-symbols the path taken by the prominence centre of mass $(\bar{x}_p(t),\bar{y}_p(t))$ up to the presently shown time. The color scale used for the cross-symbols uses time to mark earlier positions red, up to white for the present frame. This latter clearly shows that during the quiescent phase, the prominence moves slightly down and left. This motion speeds up slightly later on, to ultimately sway back to the right also in response to the coronal rain event at the prominence top. We quantified the first leftward speed as determined from the $\bar{x}_p(t)$ variation between $t=5.5-6.5$ hours to $-1869.19 \,\mathrm{km}\,\mathrm{hr}^{-1}$. The speed for the rightward return horizontal motion in time interval $t=8-9$ hours goes up to $3030.72\, \mathrm{km}\,\mathrm{hr}^{-1}$. For both the left and rightward sway, this gets combined with a vertical downward speed of $-1933.5 \, \mathrm{km}\,\mathrm{hr}^{-1}$ and $-1973 \, \mathrm{km}\,\mathrm{hr}^{-1}$, respectively. The pressure distributions clearly show that although the PCTR and TR are joined, the delicate force balance between pressure gradient, Lorentz forces and gravity still allows the prominence to be identified as a tear-drop shaped structure (in this cross-sectional view, remembering the setup to be 2.5 dimensional). We emphasize that these bodily movements of the prominence merely reflect the continued mass increase of the funnel prominence, through ongoing evaporation-condensation processes. Therefore, the whole structure moves down and further indents mass-loaded field lines, which are frozen-in for our MHD simulation. As we will quantify further on, the total prominence mass per unit length in the ignored direction is of the order $1-2 \, \times 10^4 \,{\mathrm{kg}}\,{\mathrm{m}}^{-1}$, so the overall downward movement represents a mass drainage of order ${\cal{O}}(4\times 10^{10}) \,{\mathrm{kg}} \,{\mathrm{hr}}^{-1}$, which matches perfectly with the mass drainage rate inferred for funnel prominences~\citep{liuetal}.

\subsection{Fluxrope formation and field evolution}

\begin{figure}
\includegraphics[width=\textwidth]{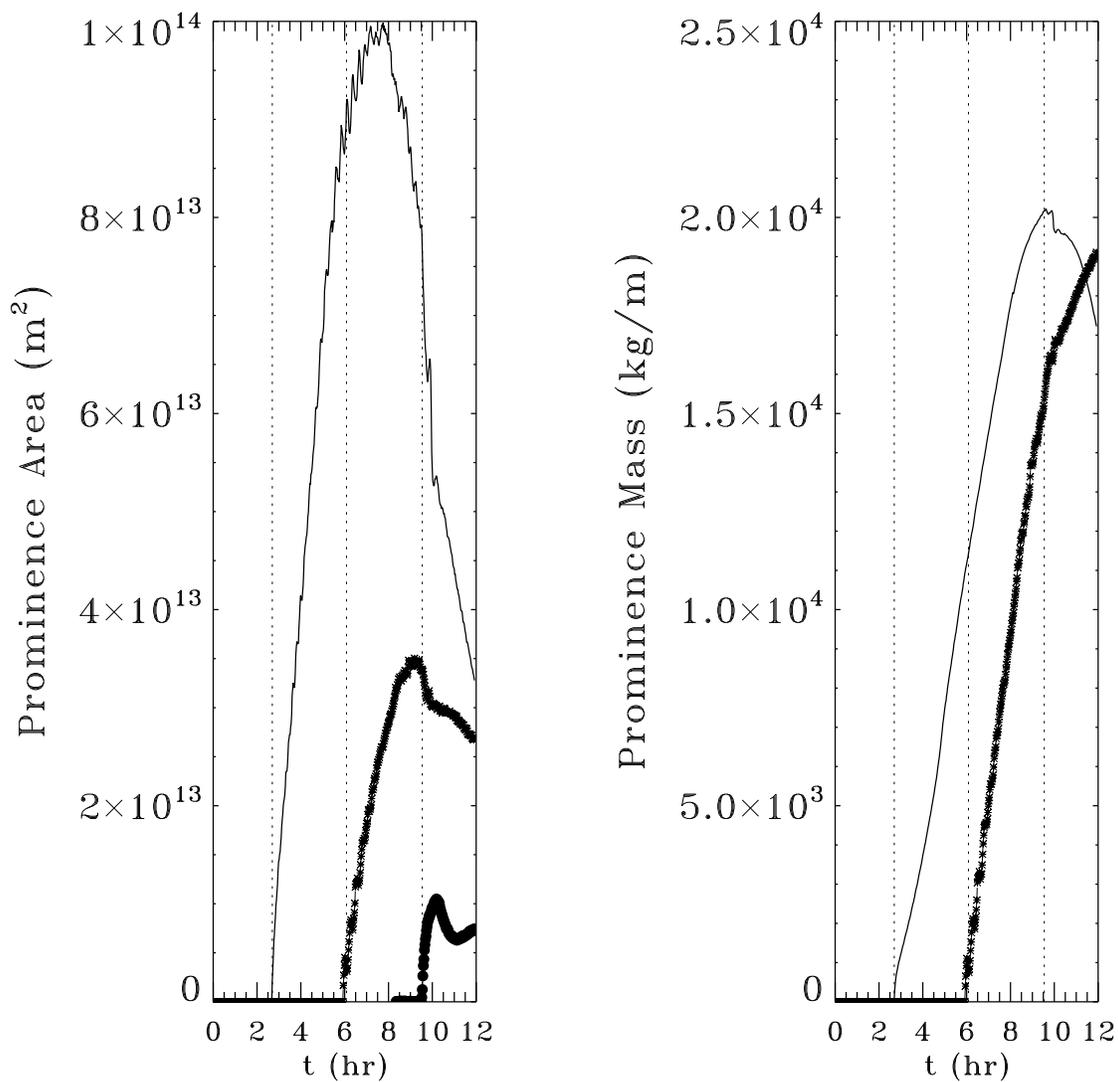}
\caption{
Using the masks for identifying prominence and core prominence plasma, the quantification of its area (left) and mass accumulated per unit length in the ignored dimension (right). Solid line for the prominence, connected asterisk symbols for the core prominence region. In the left panel, filled circles are used for quantifying the area evolution of the fluxrope, which only forms from about $t\approx 9.5$ hours.
}
\label{fsome}
\end{figure}

\begin{figure}
\centering
\includegraphics[width=0.49\textwidth]{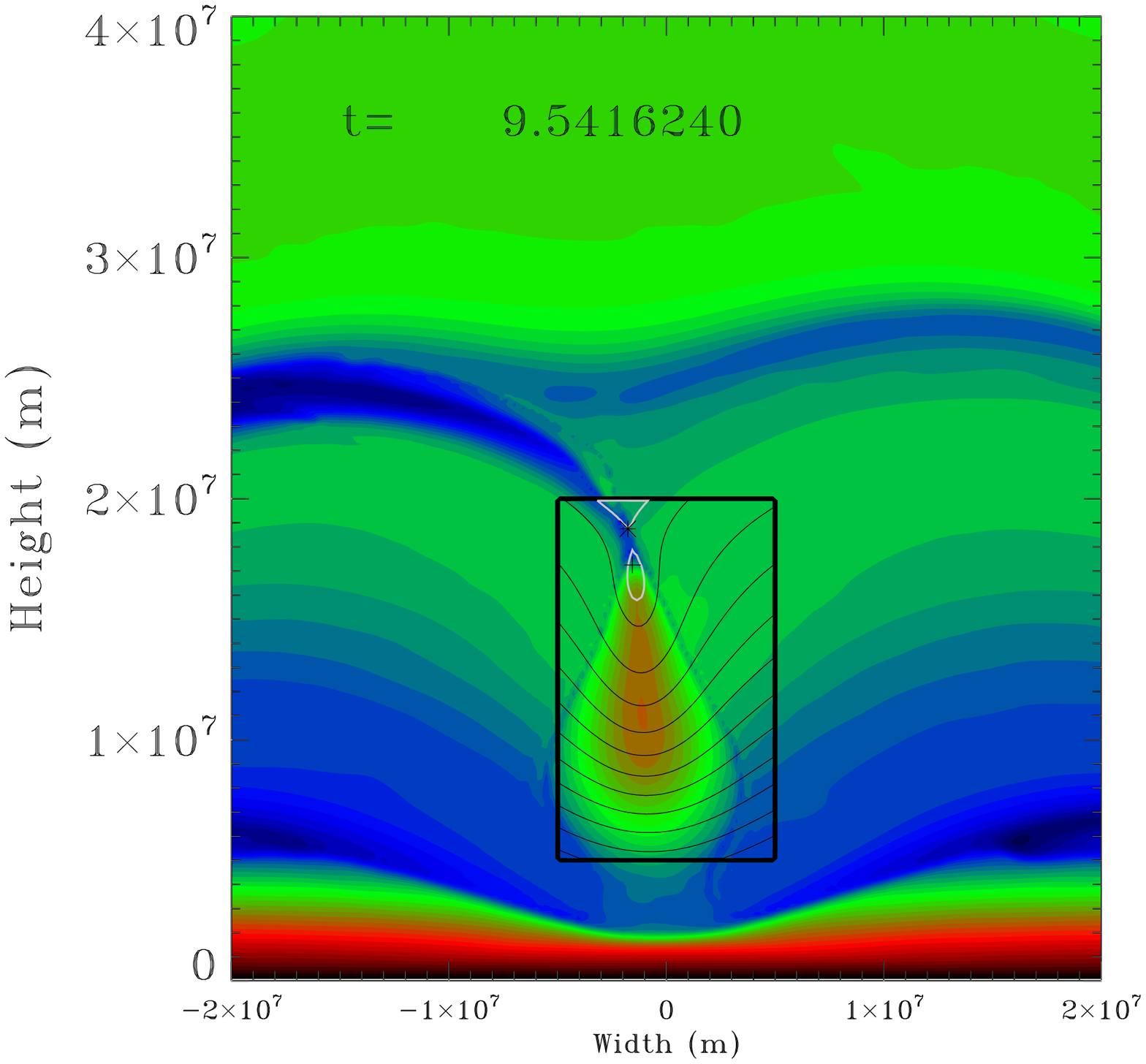}
\includegraphics[width=0.49\textwidth]{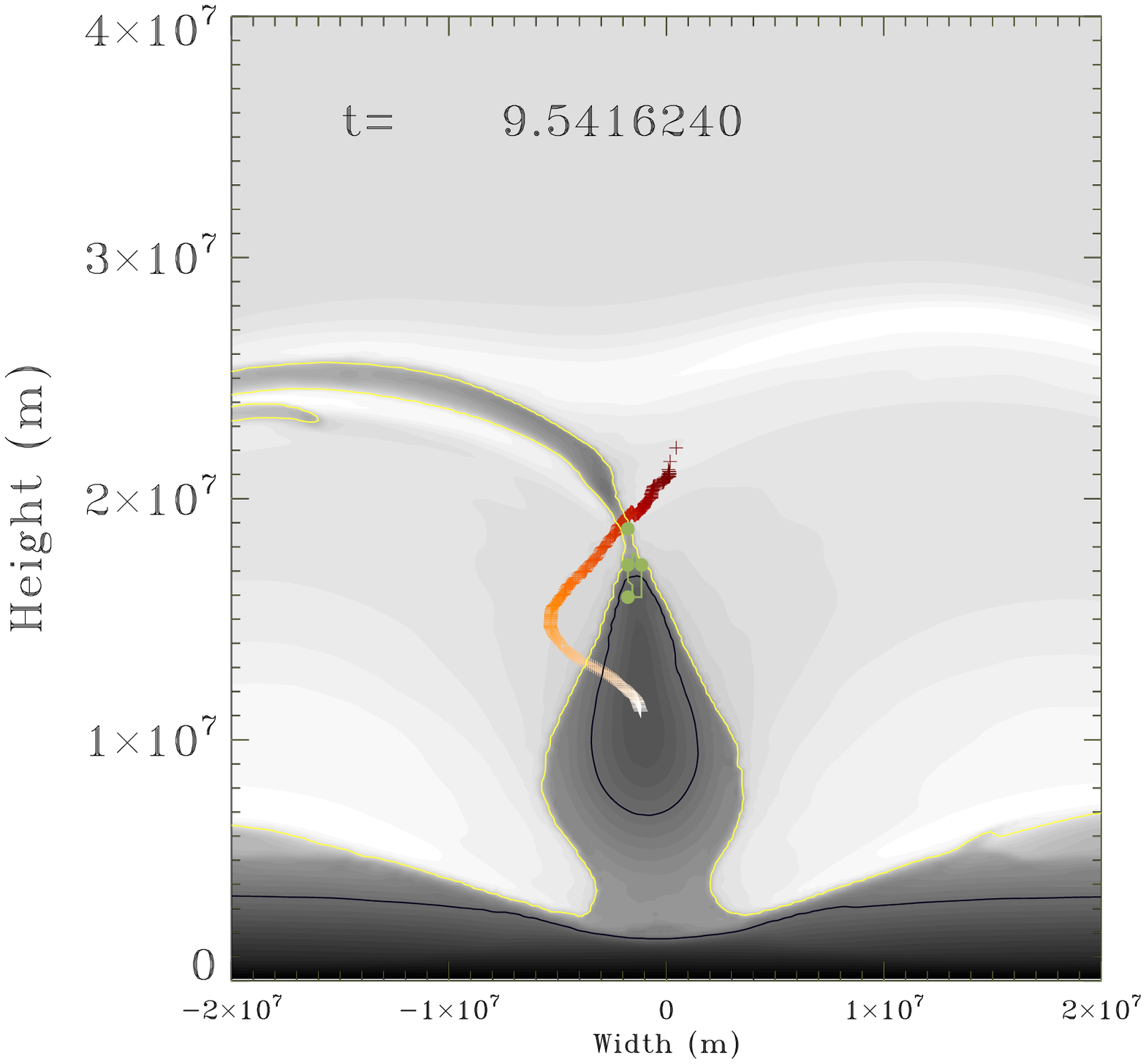}
\includegraphics[width=0.49\textwidth]{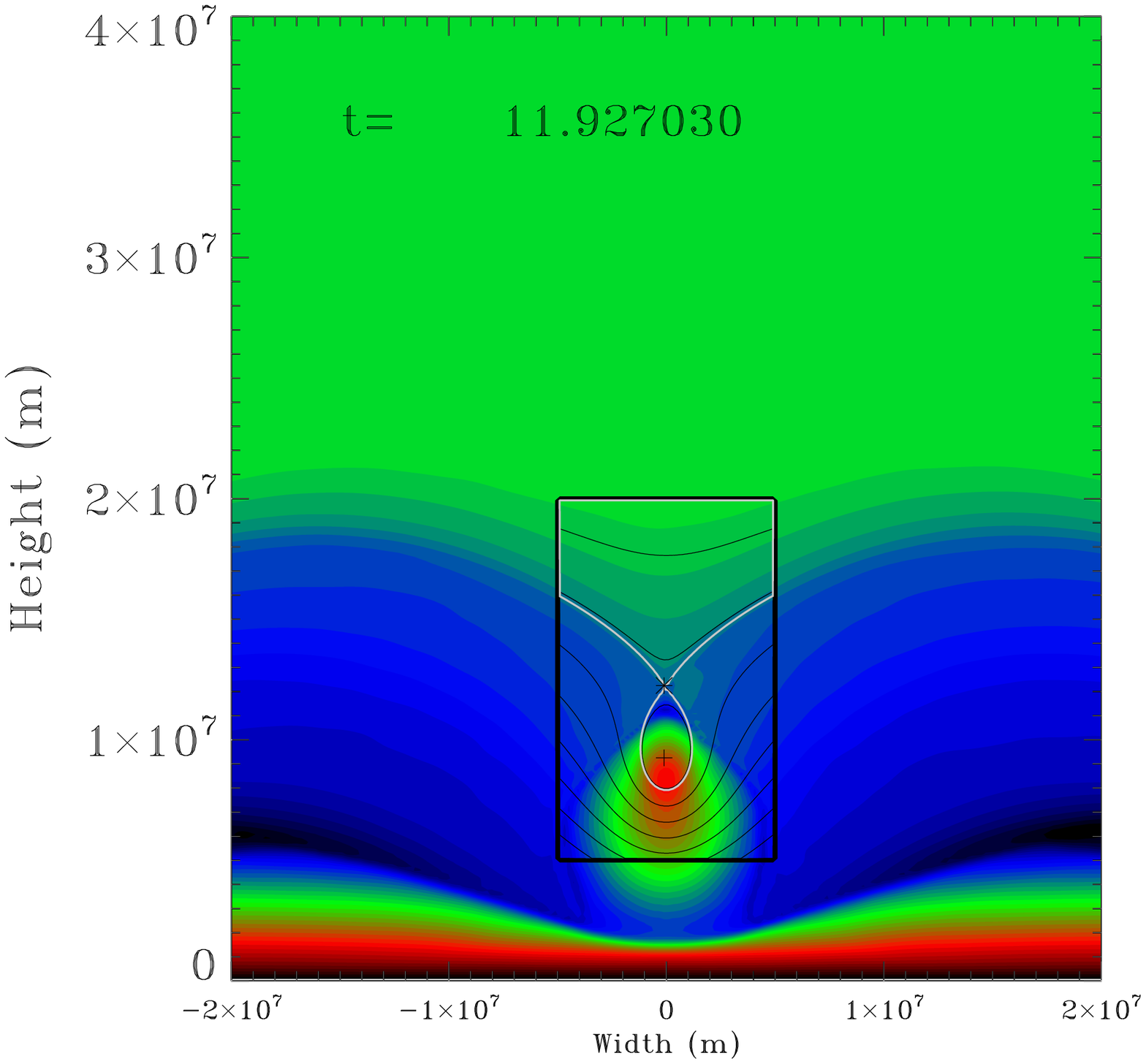}
\includegraphics[width=0.49\textwidth]{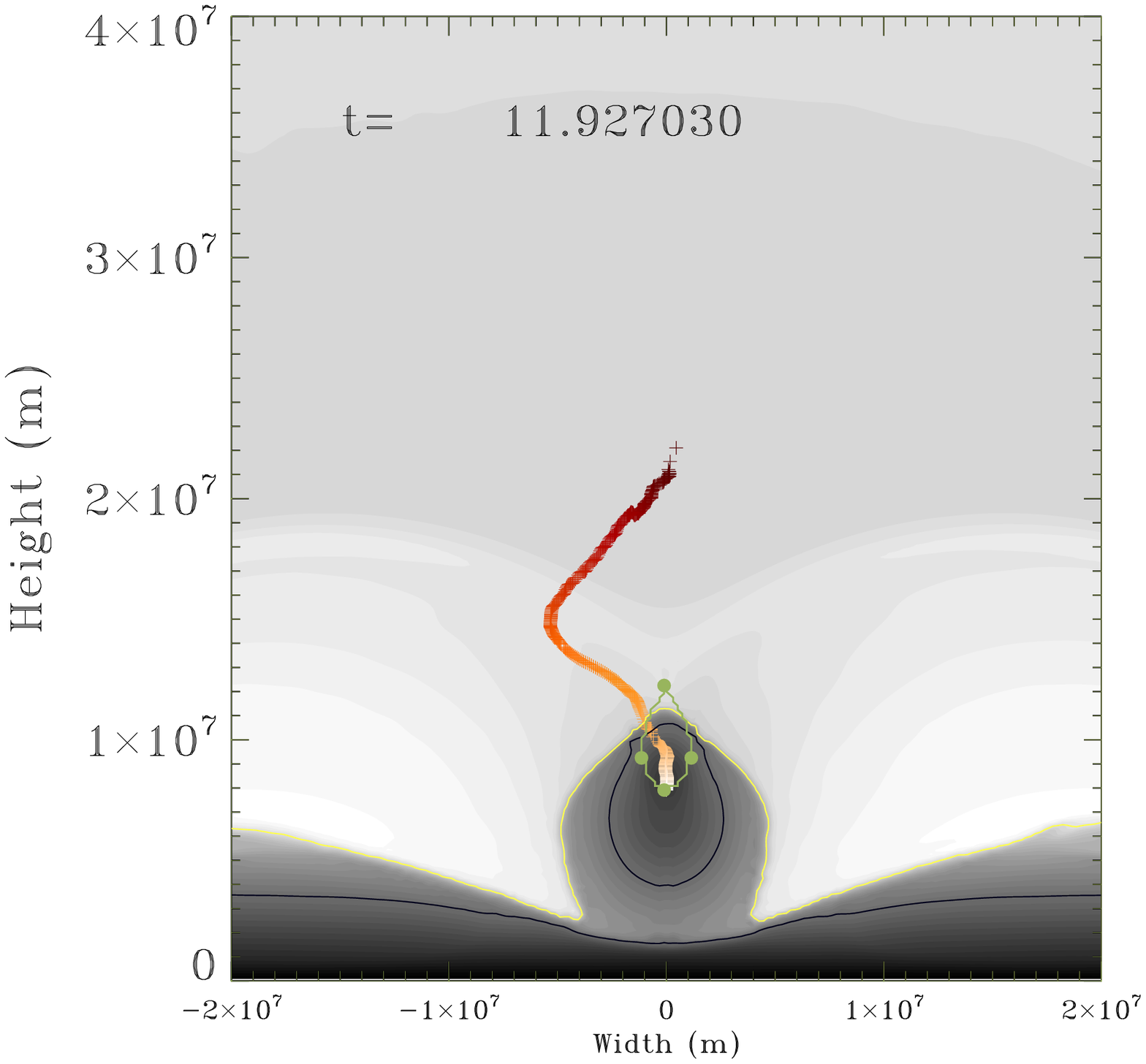}
\caption{
As in Figure~\ref{fevolvea}, for two times selected to show the fluxrope formation (top) and further evolution (bottom). The left pressure panels now also show the poloidal field lines as isocontours of the $z$-component of the vector potential, in a central area. The seperatrix contour is shown in grey, and the $X$- and $O$-point location are shown with symbols (see text). In the right density panels, we also show the fluxrope contour in green, and its extremal positions in width and height (filled circles), used to quantify the fluxrope evolution.
(A color version of this figure is available in the online journal.)
}
\label{fevolveb}
\end{figure}

At about $t\approx 9.5$ hours, the increased mass of the prominence together with the still ongoing dynamic coronal rain at its top, suddenly triggers a (necessarily numerical) reconnection event. The accumulated mass evolution, as well as the prominence area evolution can be quantified using the masks introduced earlier. This is done in Figure~\ref{fsome}, where the solid line uses the prominence mask, and the asterisk symbols use the core prominence mask, with the latter only starting from about 6 hours into the simulation. The prominence mass (right panel) clearly increases up to time $t\approx 9.4$ hours and becomes of order $1-2 \, \times 10^4 \,{\mathrm{kg}}\,{\mathrm{m}}^{-1}$, while the core systematically lags the total mass by $0.4-0.7 \, \times 10^4 \,{\mathrm{kg}}\,{\mathrm{m}}^{-1}$. Taken together with the areal quantification in the left panel, this shows that from $t\approx 6$ hours onwards, most prominence mass gets trapped within a central dense core, at all times about an order of magnitude more compact in size. This localized weigth enforces the locally upwardly bent fieldlines to dip even further, and when a kind of rebound effect occurs in the spill-over effect at the prominence tip, the field locally reconnects and introduces a magnetic island with an $X$-point on top in the projected poloidal field lines. In Figure~\ref{fevolveb}, the same zoomed views as used in Figure~\ref{fevolvea} are now augmented with local poloidal field line views, as overlaid on the central region of the pressure maps at left. At the time shown in the top panels ($t=9.54$ hours), the reconnection has just happened, and the grey contour corresponds to the separatrix. This seperatrix is here determined as the contour for the $z$-component of the magnetic vector potential that passes through the location of the local minimum in poloidal field magnitude. This minimum is indicated at left with an asterisk symbol, while a cross indicates the location of the local maximum in $B_z$. The latter identifies nicely the flux rope axis. Using the seperatrix contour, the horizontal width and vertical height of the flux rope cross-section can be detected automatically, and this is indicated on the right panels with the green contour and dot symbols (which mark the horizontal and vertical extremal locations). This recipe for determining the instantaneous area and location of the flux rope works well for all further simulated times, with an example shown at $t=11.92$ hours, at the end of our simulation in the lower panels of Figure~\ref{fevolveb}. It is seen there that the flux rope overlaps with the upper part of the prominence, and contains a part of the core prominence (within the black contour in the right panels), and we can use its automated detection to similarly quantify the area evolution of the flux rope. This is actually also shown in Figure~\ref{fsome}, with circle symbols in the left panel. It can be seen that the flux rope is present for the entire 2.5 hour period, with cross-sectional area of about $0.8 \times 10^{13} \,\mathrm{m}^2$. As shown later, the width settles at about $2 \times 10^6 {\mathrm{m}}$ while its height becomes of order 4 Mm. The moment of flux rope formation is also clearly detected in several of the time traces discussed before, especially in the prominence average plasma beta in Fig.~\ref{fdriv}, where it represents a clear decrease, along with an increased magnetic energy content for the core prominence plasma especially (Fig.~\ref{fenergy}). The dynamics at the prominence top both involve a loss of matter and prominence cross-sectional area (see Figs.~\ref{fsome}), and a magnetic topological change that locally raises the current and Lorentz force influence. Both add to the altered plasma beta conditions. As an aside, Fig.~\ref{fsome} gives seemingly contradictory trends for area and prominence mass when comparing total to core plasma evolutions in the final stages, but this is due to the geometric clipping below fixed heights as used in the masks discussed (needed because the prominence seemingly connects to the upper chromosphere).

\begin{figure}
\centering
\includegraphics[width=5.8in]{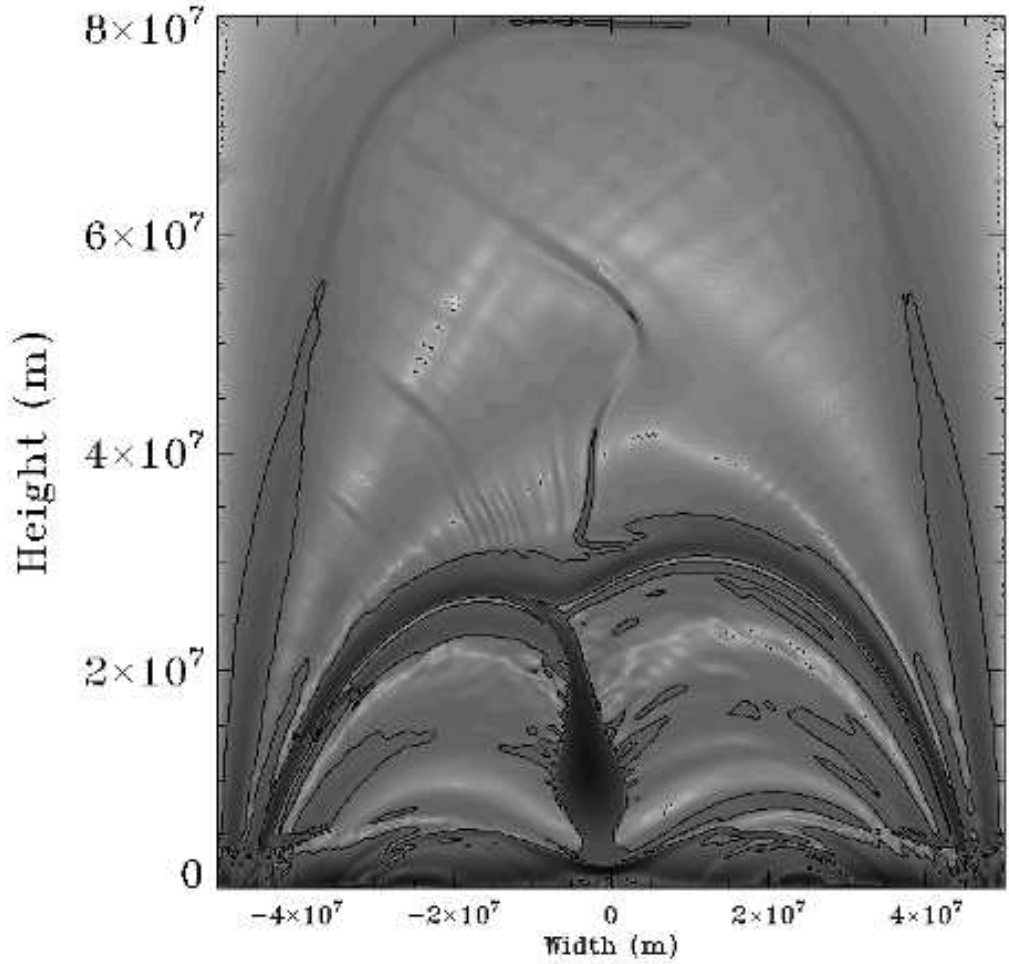}
\caption{
An arbitrarily stretched greyscale map for the magnitude of the Lorentz force, shown on the whole domain, at time $t=8.66$ hours. The fluxrope has not yet formed, but the coronal rain, i.e. prominence plasma overspilling to the left arcade part, has launched strong rebound wave trains, seen here at heights $y\approx 3.5 \,\times 10^{7}\,\mathrm{m}$, causing strong undulatory wave patterns in the overlying arcade.
}
\label{fwaves}
\end{figure}

We stressed that the fluxrope formation is triggered by a coronal rain event, where the prominence body spills over the left half of the arcade. That this is indeed a dynamic event, with enhanced Lorentz forces acting as restoring force and ultimately allowing the (numerical) reconnection, is made visual by an instantaneous view of the Lorentz force magnitude across the entire domain. This is shown in Fig.~\ref{fwaves}, where we used an arbitrary, logarithmically stretched color scale to enhance both small (linear) as well as large amplitude variations at time $t=8.66$ hours. An animated view for the entire simulation is provided as online material. This full domain view shows how adjacent to the prominence feature, but also in the overarching arcade part, clear signatures of wave propagations, reflections and interference patterns can be detected. Animated views allow to trace a particularly interesting part of the evolution where the prominence matter overspills repeatedly, and the interaction of the falling blobs with the transition region plasma causes rebound wave patterns, leading to clear coherent oscillations of the overarching field lines above the prominence. These oscillations travel from left to right (and back), and at the snapshot shown in Fig.~\ref{fwaves} correspond to the almost vertically oriented wavetrain patterns seen above the prominence structure: the leading front is now at about $x\approx 0$ and stretches across heights $y\approx 3-4\times 10^7 {\mathrm{m}}$. In animated views of the field line structure as in Figs.~\ref{fevolve1}-\ref{fevolve2}, one can easily detect the corresponding wave motions. Their speeds range around $115.5 \,\mathrm{km}\,\mathrm{s}^{-1}$, estimated from approximating the fieldline by an ellipse and noting that in about 6 time units the front propagates from transition region to the middle of the arcade. These strong wave motions (in fact, the entire simulation shows a lot of linear wave dynamics, but we deliberately focus on more nonlinear features here) occur up to and beyond the period of fluxrope formation (at $t\approx 9.5$ hours). Their repeated passage, while the prominence gets increasingly weighed down by its core, certainly influence the local accuracy of our simulation (the tip of the prominence is a region of high gradients in both temporal and spatial sense) and facilitates numerical reconnection there. Still, we are tempted to interpet this as a physically realizable change in magnetic topology, permitted when finite resistivity would be included. Revisiting this part of the evolution with (local or otherwise anomalously motivated) finite resistivity is left to future work. For the purpose of this paper, we now further argue how such fluxrope formation within a mostly bipolar arcade system would evolve.

\begin{figure}
\centering
\includegraphics[width=\textwidth]{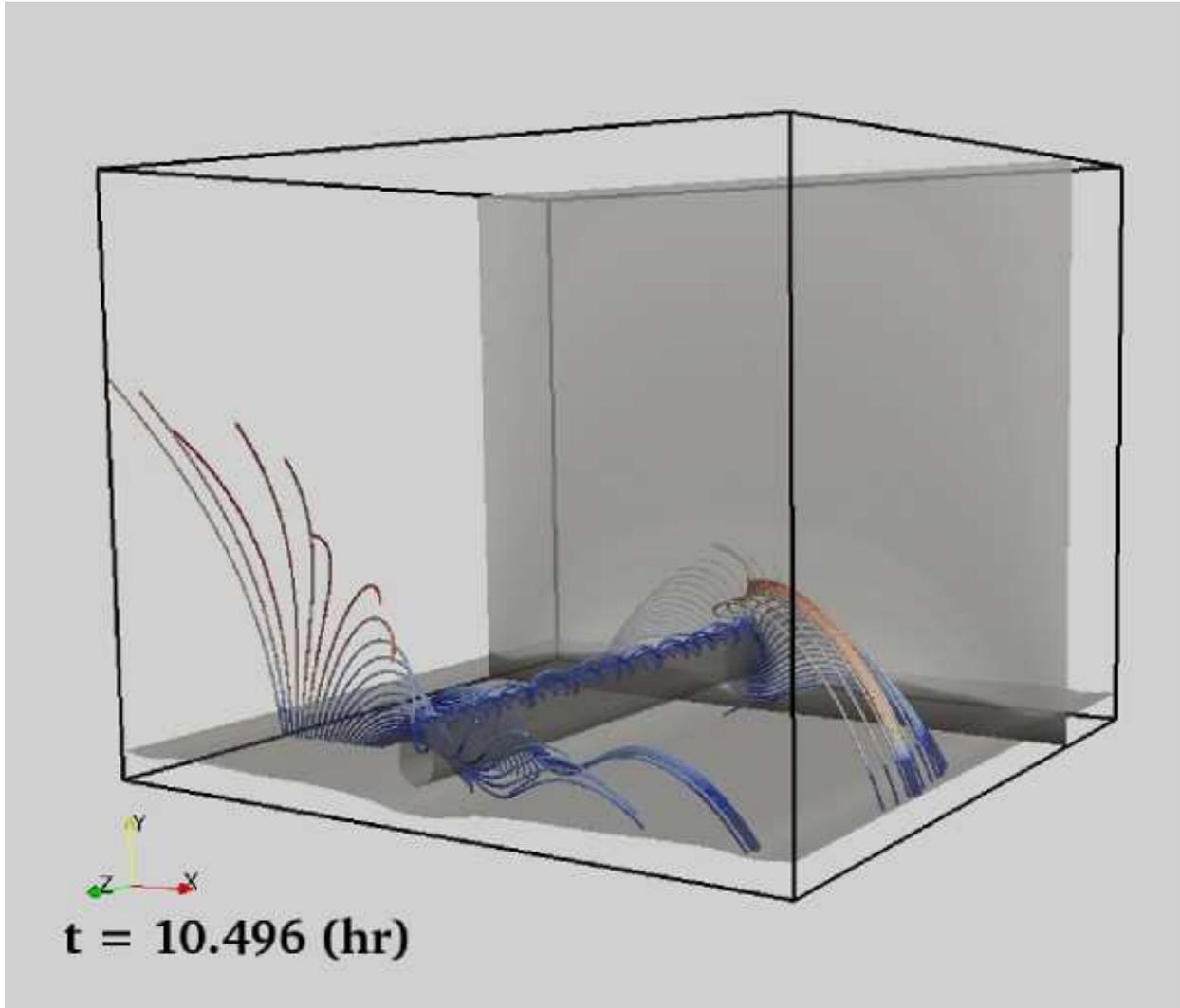}
\caption{
A 3D view on the later stages of our simulation where a fluxrope has formed and traps the top part of the prominence plasma. Shown are selected field lines colored by temperature (as in Figs.~\ref{fevolve1}-\ref{fevolve2}), the vertical cross-section shows the density, and a density isosurface at $2.341 \times 10^{-10}\,\mathrm{kg}\,\mathrm{m}^{-3}$ identifies the core prominence region (and a similar isosurface underneath the TR). The fluxrope has clearly twisted fieldlines passing through the core prominence.
(A color version of this figure is available in the online journal.)
}
\label{f3dview}
\end{figure}

To do this, we first provide a three-dimensional view on our 2.5D simulation in Fig.~\ref{f3dview}. We generate this by simply repeating the data in the ignored $z$-direction, taking the $z$ extent 100 Mm wide (identical to the $x$-range). At time $t=10.496$ hours, the field structure is shown by selected field lines, colored by temperature in a similar fashion to the 2D views in Figs.~\ref{fevolve1}-\ref{fevolve2}. In the vertical cutting plane, we similarly use grayscale for density, and augment it with two isosurfaces of the density that necessarily are invariant in the $z$-direction. These two isosurfaces are obtained for fixed density value $\rho=100 \times \rho_{\mathrm{unit}}=2.341 \times 10^{-10} \,\mathrm{kg}\,\mathrm{m}^{-3}$, which was used earlier to detect the core prominence plasma region. Therefore, we obtain a nearly horizontal isosurface in the upper chromosphere, and a clear tube structure with an egg-shaped cross-section marking the core prominence. The flux rope is seen here by the twisted fieldlines going through the upper part of this latter isosurface. A fair amount of twist is present in this fluxrope, but because we simulate in 2.5D, the flux rope axis can never deform from a straight line oriented along the $z$-direction.

\begin{figure}
\centering
\includegraphics[width=5.8in]{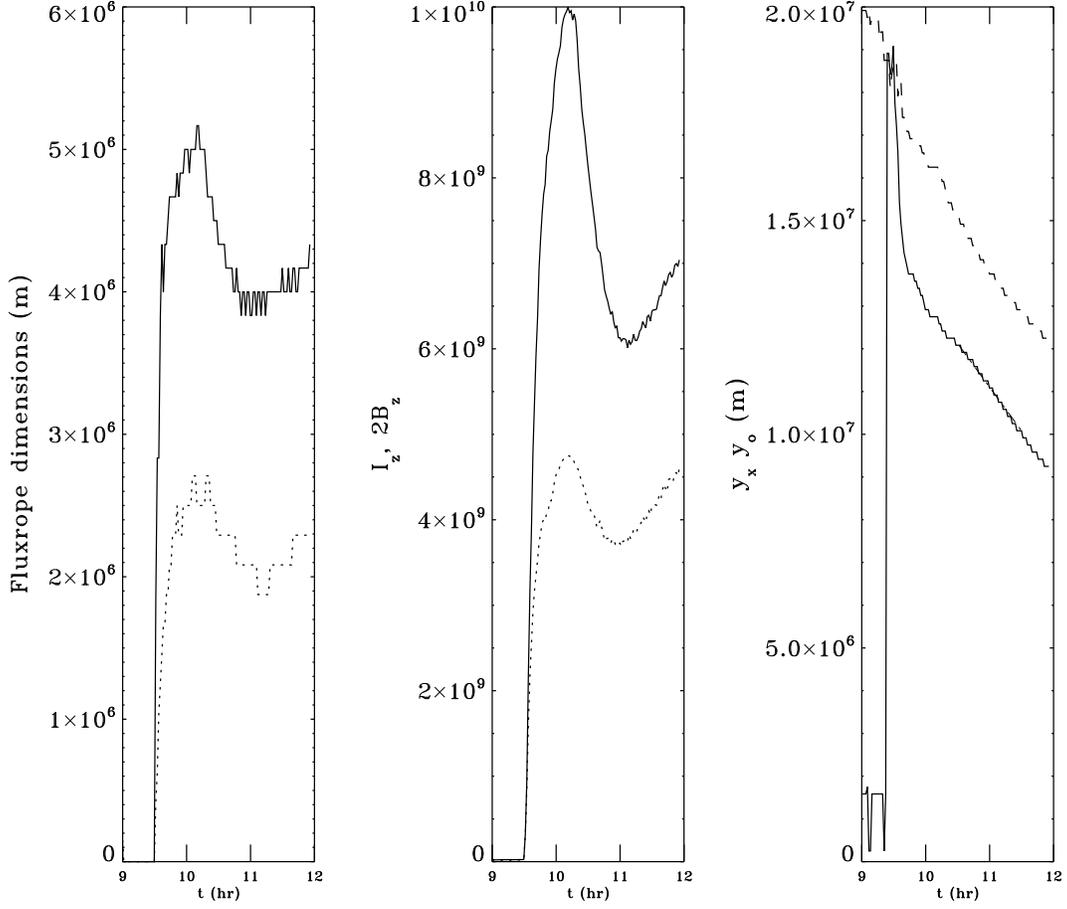}
\caption{
Using the poloidal field topology to identify the fluxrope region, we show at left the flux rope width (dotted) and height (solid) evolution; in the middle the fluxrope cross-section integrated current (dashed) and $z$-component of the magnetic field (solid) (where we used mksA units for both); at right the heights of the $X$-point (dashed) and $O$-point (solid).
}
\label{finit}
\end{figure}

To discuss the fate of this fluxrope, once formed, we present in Fig.~\ref{finit} several time histories of its most distinct properties. As explained for Fig.~\ref{fevolveb}, we can at all times quantify the fluxrope area through its bounding magnetic potential contour. The extremal $x$- and $y$-coordinates on this contour quantify the width and height of the fluxrope, as plotted at far left in Fig.~\ref{finit}. At all times, the fluxrope is twice as long as it is wide, and its area was shown in Fig.~\ref{fsome} in comparison to the prominence and core prominence cross-sectional evolution. With our highest resolution being $\Delta x=0.20833$ Mm, the width is resolved with over 10 grid points, enough to meaningfully quantify internal properties. The rightmost panel from Fig.~\ref{finit} shows the vertical ($y$) position of the $X$- and $O$-points as found in the magnetic potential views, and they are seen to show a sustained seperation of about 3 Mm, and a consistent downward motion with $v_y=-2045 \, {\mathrm{km}}\,{\mathrm{hr}}^{-1}$ is present. This is of the same order of magnitude as the downward speeds inferred earlier from the movements of the centre of mass for the prominence plasma, which were mentioned for earlier phases of the evolution. Although this thereby matches again with the previously quoted values for mass drainage down to the chromosphere, we rather speculate in what follows that this part of our 2.5D simulation is unrealistic, since a more likely fate for the twisted fluxrope is a violent eruption by kink instability. If that would happen, a fact that can only be tested with true 3D follow-up simulations, a fraction of the prominence trapped in the fluxrope proper would be ejected in a coronal mass ejection event. To make this speculation plausible, the middle panel of Fig.~\ref{finit} quantifies the instantaneous flux-rope area integrated $z$-current (the dotted line, being $\iint J_z \, dx\,dy$ over the fluxrope only), as well as the integral $2\iint B_z \,dx\,dy$. When plotting these integrals, we used mksA units. We now note~\citep{book1} that the Kruskal-Shafranov limit for external kink mode stability of a current carrying plasma column with area element $dA$ embedded in vacuum renders stability as long as
\begin{equation}
I_z=\iint J_z \, dA < \frac{2}{\mu_0 R_0 }\iint B_z \,dA  \,.\label{ks}
\end{equation}
In this expression, the as yet unaccounted factor $\mu_0 R_0$ in the denominator contains the assumed `length' in the ignored direction, when a cylindrical column of radius $a$ and length $L=2\pi R_0$ has inverse aspect ratio $\epsilon=a/R_0$ mimicking a `straight' tokamak configuration. The factor 2 in $L$ may not be needed when only half-wavelength modes are possible, as in our translationally invariant situation. In any case, a typical prominence length easily reaches $L=10^8 \mathrm{m}$ as taken for our mock-up 3D view in Fig.~\ref{f3dview}. With the observed area from Fig.~\ref{fsome} and sizes in Fig.~\ref{finit}, we have a fluxrope radius $a\leq 3 \times 10^6 \,\mathrm{m}$, making its inverse aspect ratio $\epsilon \approx {\cal{O}}(0.1)$. In the mksA units adopted throughout this paper, Eq.~\ref{ks} must then account for a factor $\mu_0 R_0$ of order unity, in combination with the plotted values in Fig.~\ref{finit}, middle panel, for the integrated quantities. Since the integrals are of the same order of magnitude, a possible route to instability via kink deformation results when the length of the prominence in $z$ increases as time progresses, and thereby leads to a sudden violation of the Kruskal-Shafranov stability limit. Although this argument needs full 3D runs for its verification, and the conditions in the fluxrope surroundings are not those for a plasma-vacuum setup, the accumulated experimental knowledge on tokamak plasma discharge behavior argues in favor of this route to (partial) prominence ejection. Observationally, this could also be verified by monitoring the evolution of funnel prominences over similar hourlong time periods, quantifying the overall dimensional changes with time, and looking for signatures of kink unstable evolutions.

\section{CONCLUSIONS AND OUTLOOK}\label{conclusion}

We presented a 2.5D simulation where prominence formation and evolution could be studied, as initiated by thermal instability from chromospheric evaporation-condensation. Our approach follows~\cite{Xia2012ApJ}, generalizing its findings to a more complex field topology of a sheared quadrupolar arcade system. The phase of gradual evolution to catastrophic condensation, and the first few hours of quiescent evolution with continued mass accumulation confirms the earlier findings on how an overall force-balanced MHD configuration gets established while the prominence grows. New insights are obtained for the more dynamic phases occuring several hours after the first condensation: the prominence shows systematic bodily motion, and can ultimately spill over its bipolar, dipped fieldlines. 
The simulation showed how the resulting coronal rain impacts can set off wave trains, inducing strong wave undulations in the overlying arcade parts. We identified a likely, novel route to coronal mass ejecta where the upper part of a hedgerow prominence could be lost by kink-unstable fluxrope evolution. Our translationally invariant model could not account for the ejection itself, and full 3D simulations are required to verify our speculations on its liability to kink deformation as soon as the length in the currently ignored dimension exceeds the treshold. Similarly, we provided arguments in favor of fluxrope formation by the increased mass accumulation enhancing the field line dips (although the needed resistive aspects are only approximately treated here), concurrent with coronal rain and wave dynamics. Our prominences were chosen to mimick realistic conditions in a quiet sun arcade configuration, and the mass drainage and overall morphological appearance provide theoretical confirmation for recent observational findings on the funnel prominence category~\citep{liuetal}. 

Aspects that need further modeling efforts along these lines include the following. Firstly, a more parametric survey of the prominence formation process is called for. The most important parameters relate to the magnetic field strength (and topology), in combination with the adopted heating prescription. Such parametric study has been initiated~\citep{fang14} for a bipolar arcade setup similar to the work by~\cite{Xia2012ApJ}, and they point out that similar energy inputs lead to similar prominence growth and that the stronger magnetic fields may resist or delay the field line bending as influenced by how fast matter accumulates, and rather lead to drainage events. These findings need to be revisited in the current quadrupolar arcade setup. Additionally, in view of the early kinematic approach by~\cite{ivanov}, it is of interest to see how time-varying bottom magnetic field conditions may introduce also the alternative means for prominence matter accumulation, by lifting of lower-lying chromospheric matter upwards due to reconnection. The current 2.5D simulation shows as yet no evidence for observed small-scale internal dynamics in the form of Rayleigh-Taylor fingering. This is likely a combination of numerical resolution and accuracy, and the specific model parameters (especially the parametrized heating) influencing the obtained density contrast at the PCTR, as well as the magnetic field strength and its variation. The restricted dimensionality also suppresses any modes which require finite wavelength in our ignored direction, so potentially the step to 3D simulations may already alleviate this aspect. In 3D, the parametrized heating could be distributed on both arcade endpoints in a fair variety of ways, and different types of dipped arcade configurations may favor coronal rain versus more large-scale prominence formation. The presented evolution can also be used for generating synthetic observations, in a spectropolarimetric sense, of the prominence formation process. This will certainly aid in confronting the wealth of observational knowledge, and can try to distinguish whether other than evaporation-condensation scenarios are at play in formation (like levitation or injection), and at what relative frequency. Another aspect for follow-up analysis is the omnipresent wave dynamics, also in the quiescent phase of our simulation, which can make contact with modern coronal seismological studies for prominences. Our model provides strong theoretical support to the thermal instability pathway, which has been studied mainly in restricted rigid field-aligned models. Physics aspects requiring further attention in our approach are: (1) the role of finite resistivity and reconnection when the fluxrope gets formed; (2) improving the ad-hoc parametrization of the added (and background) heating, by e.g. using wave-propagation and dissipation prescriptions as used in the most recent models for the global corona to heliosphere by~\cite{awsom}; and (3) allowing for partial ionization effects, as these are known to be important in prominences, and in turn alter their liability to Rayleigh-Taylor mode development~\citep{komenko}. Another line of research needs to investigate the same processes in MHD-stable 3D fluxrope settings. Therefore, true finite-beta flux rope formation in the presence of gravity~\citep{xia14}, extended to full thermodynamics with chromosphere-transition region-coronal layering~\citep{xia14b}, will need to demonstrate how cavities can result from in-situ condensation in fluxropes. 

\acknowledgments
This research was supported by projects GOA/2015-014 (2014-2018) (KU Leuven), 
FWO Pegasus funding, and the Interuniversity Attraction Poles Programme 
initiated by the Belgian Science Policy Office (IAP P7/08 CHARM). The 
simulations used the VSC (flemish supercomputer center) funded by the Hercules 
foundation and the Flemish government. 

\bibliographystyle{apj}

\clearpage

\clearpage

\clearpage

\clearpage

\clearpage

\clearpage

\clearpage

\clearpage

\clearpage

\clearpage

\clearpage
\end{document}